\documentclass[aps,prd,showpacs,twocolumn,nofootinbib,superscriptaddress,amsmath,amssymb,8pt]{revtex4-2}
\usepackage{amsmath,amssymb,amsfonts}
\usepackage{bm}
\usepackage{verbatim}
\usepackage{hyperref}
\usepackage{slashed,braket}
\usepackage{color}
\usepackage{graphicx,graphics}
\usepackage[title,titletoc]{appendix}
\usepackage{stackengine}

\newcommand{\cA}{\mathcal{A}}
\newcommand{\cD}{\mathcal{D}}
\newcommand{\cF}{\mathcal{F}}

\newcommand{\Tr}{\operatorname{Tr}}

\newcommand{\tr}{\operatorname{tr}}

\newcommand{\D}{\Delta}

\newcommand{\red}[1]{{#1}}
\newcommand{\mz}[1]{{#1}}
\newcommand{\mzz}[1]{{#1}}
\newcommand{\mzo}[1]{{#1}}

\usepackage{definitions}%

\begin{document}

\title{Effect of interactions on the topological expression for the  chiral separation effect}

\author{M.A. Zubkov}
\email{mikhailzu@ariel.ac.il}
\affiliation{Physics Department, Ariel University, Ariel 40700, Israel}

\author{Ruslan A.~Abramchuk}
\email{abramchuk@phystech.edu}
\affiliation{Physics Department, Ariel University, Ariel 40700, Israel}

\date{\today}

\begin{abstract}
    In the absence of interactions the conductivity of chiral separation effect (CSE) in the system of massless fermions is given by topological expression. Interactions might change the pattern drastically. However, we prove that the CSE conductivity is still given by the topological invariant composed of the Green functions at zero temperature as long as the chiral symmetry is present, \mzz{and if the renormalized axial current is considered.} This allows to predict its appearance with the standard value of conductivity per Dirac fermion $\sigma_{CSE} = \frac{1}{2 \pi^2}$ in quark - gluon  matter at $T = 0$ and sufficiently large baryon chemical potential, in the hypothetical phase with restored chiral symmetry and without color superconductivity. This phase may be realized inside the neutron stars. We also argue that the same topological expression for the CSE may be observed in Weyl semimetals, which realize the system of interacting relativistic fermions in solid state systems. In order to estimate the non - perturbative corrections to $\sigma_{CSE}$ within QCD at finite temperatures  we apply method of field correlators developed by Yu.A.Simonov. As expected, above the deconfinement crossover the topological expression is approached within the quark - gluon plasma phase, when the quark chemical potential is sufficiently large. However, we observe that this occurs only when quark chemical potential is much larger than  the thermal (Debye) mass. This range of parameters appears to be far out of the region accessible at the modern colliders.
\end{abstract}

\maketitle
\tableofcontents
\section{Introduction}
\label{SectIntro}

The non - dissipative transport effects appear both in condensed matter physics and in the high energy physics \cite{Metl,Kharzeev:2013ffa,Kharzeev:2015znc,Kharzeev:2009mf,ref:HIC,9,Landsteiner:2012kd,semimetal_effects6,Gorbar:2015wya,Miransky:2015ava,Valgushev:2015pjn,Buividovich:2015ara,Buividovich:2014dha,Buividovich:2013hza}.These effects represent an important probe of the corresponding systems because of their topological nature. The corresponding conductivities, as expected, in many cases are represented by the topological invariants robust to any smooth modifications of the systems including switching on interactions. As a result the  strong interactions, which cannot be taken into account using direct calculations, do not have any effect on these quantities. An example is given by QCD at finite baryon chemical potential. Here the lattice numerical simulation cannot be applied, while the non - perturbative effects still remain essential. In particular, the appearance of color superconductivity predicted with the aid of perturbative QCD  is questionable because of the non - perturbative nature of QCD even at large $\mu_B$.

\begin{figure}[h]
	\centering  %
	\includegraphics[width=0.9\linewidth]{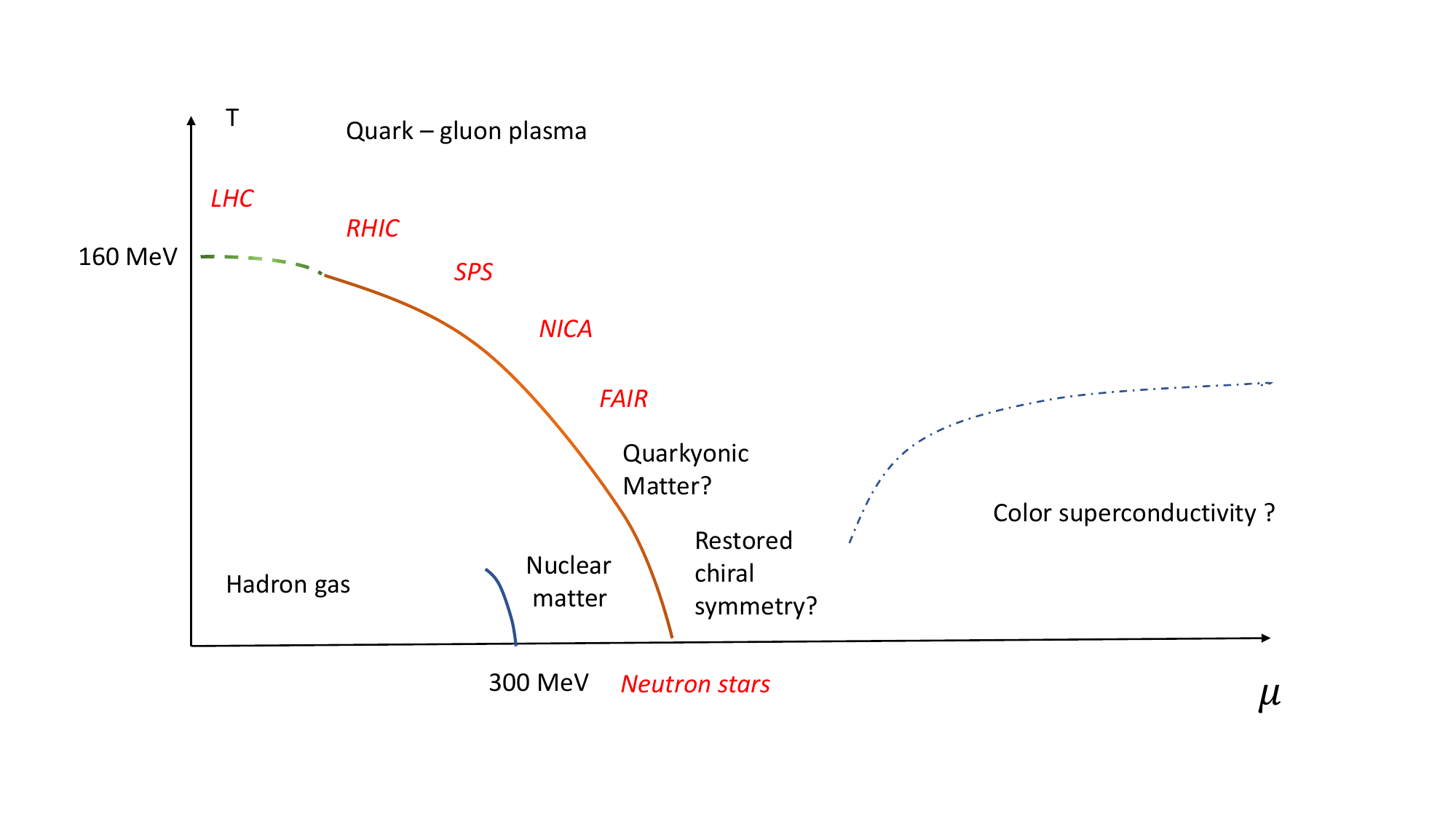}  %
	\caption{We represent here the sketch of the QCD phase diagram in the plane temperature - quark chemical potential. The deconfinement crossover is represented by the dashed line. Here we have the data obtained using lattice simulations. Above the dashed line there is the quark - gluon plasma phase with restored chiral symmetry and deconfinement. At larger values of chemical potential the crossover, presumably, is changed by the true phase transition of the first order. Its nature for small temperature is not yet well established. We may, actually, have here separate lines of deconfinement phase transition and the chiral symmetry restoration transition. Left to this transition (these transitions) there may be the quarkyonic phase with co - existing baryons and quarks, and the deconfining phase with restored chiral symmetry. At extremely large values of quark chemical potential several color superconductor phases might appear. The region in the lower left corner of the phase diagram is separated by the first order transition line to the phase of hadronic gas and nuclear matter. Modern colliders (LHC, RHIC, SPS, NICA, FAIR) will be able to probe the regions of parameters of the phase diagram along the line of the phase transition. The interior of neutron stars represent the laboratory for probe of the domain with small temperatures and large chemical potentials. According to the common lore this region is not accessible at the present moment for the existing non - perturbative QCD calculations (see, however, \cite{Krivoruchenko2010jz}). }  %
	\label{phase_diagram}   %
\end{figure}

 The sketch of the phase diagram of QCD is represented in Fig. \ref{phase_diagram} in the plane temperature - quark chemical potential \cite{QCDphases,1,2,3,4,5,6,7,8,9,10}. Vacuum of the theory ($T=\mu=0$) is situated in the left lower corner of the diagram. The dashed line represents well - investigated deconfinement crossover. Above this crossover the quark - gluon matter is in the quark  - gluon plasma phase, which is still a strongly correlated medium, with deconfinement and restoration of chiral symmetry. The dashed line is assumed to transform into the true phase transition line. This line meets the axis $T=0$ somewhere above $\mu = 300$ MeV. However, as it was mentioned above, the region of finite $\mu$ is not accessible for lattice numerical simulations. The perturbative analytical calculations cannot describe the quark - gluon matter exhaustively. At the small values of $\mu$ certain  methods of calculations may be used based on lattice numerical simulations as those based on the expansion of considered physical quantities in powers of $\mu$. The only clear result at the right hand side of the diagram is the line separating the phase of hadronic gas from nuclear matter. Qualitatively transition to nuclear matter occurs when quark chemical potential becomes larger than the constituent quark mass. In this situation quark matter becomes as dense as it is inside the atomic nuclei. Further increase of chemical potential might lead to several  phase transitions \cite{Simonov2007jb}. There is a number of hypotheses about these transitions. The corresponding part of the solid line may represent transition to one of the phases of color superconductivity. In addition, right to this line the so - called quarkyonic phase may be situated, where the quarks co - exist with baryons. In this phase there is confinement of quarks and chiral symmetry is broken, as well as left to the transition line, but the sea of the particles inside the Fermi sphere consists of separate quarks. Another supposition is that the vertical transition line is to be separated to the two: the confinement - deconfinement transition and the chiral symmetry restoration line. As it was mentioned above, the phenomenological methods based on perturbative QCD  predict appearance of several color superconductor phases right to the vertical transition line. The lower right corner of the phase diagram is typically associated with the color - flavor locking color superconductor phase.

The complementary arena for the experimental observation of non - dissipative transport effects is represented by electronic quasiparticles in Dirac and Weyl semimetals \cite{semimetal_effects6,semimetal_effects10,semimetal_effects11,semimetal_effects12,semimetal_effects13,Zyuzin:2012tv,tewary,16}.
These materials simulate in laboratory physics of relativistic elementary particles. Interactions between them break emergent relativistic invariance, but even so these materials are an important probe of elementary particle physics with strong interactions.

The chiral separation effect (CSE) has been proposed by M.Metlitski and A.Zhitnitsky \cite{Metl}, and it is the representative of the family of non - dissipative transport phenomena. This effect results in axial current directed along external magnetic field in the presence of non - zero chemical potential. Originally this effect has been considered in the system of continuum Dirac fermions, which would be homogeneous without external magnetic field. In the chiral limit, i.e. for massless fermions the axial current in these systems is proportional to the external magnetic field strength $F_{ij}$ and to the ordinary chemical potential $\mu$:
\begin{equation}
	J_5^k = \frac{1}{4\pi^2}\epsilon^{ijk0} \mu F_{ij}\label{1}
\end{equation}
It has been proposed that this effect may be observed in the quark - gluon plasma (QGP) phase. In particular, the possibility to observe this effect experimentally during heavy ion collisions has been discussed \cite{Kharzeev:2015znc,Kharzeev:2009mf,Kharzeev:2013ffa,ref:HIC}. The fireballs appeared in the non - central collisions of heavy ions are supposed to realize the QGP phase, and are  subject to strong magnetic field \cite{QCDphases,1,2,3,4,5,6,7,8,9,10}. In the QGP phase there is no confinement of quarks, and the chiral symmetry is restored. The two colliding ions produce strong magnetic field. After the decay of the fireball, in principle, the signature of the CSE may be found in the asymmetry of created particles. The CSE may also be relevant for the description of the quark - gluon matter at the other side of the QCD phase diagram - inside the neutron stars~\cite{Cook:1993qr}. Extension of the consideration of CSE to the essentially non  - homogeneous systems has been performed in \cite{SZ2020}, where it has been shown that the non - homogeneity does not affect the CSE conductivity, which remains topological invariant proportional to the number of the species of Dirac fermions.

Relation of CSE to chiral anomaly has been considered in   many works  \cite{Zyuzin:2012tv}). The cousin of the CSE - the Chiral Magnetic Effect  (CME) \cite{Vilenkin,CME,Kharzeev:2013ffa,Kharzeev:2009pj,SonYamamoto2012} has also been conjectured to be related to chiral anomaly. It has been shown, however, that in thermal equilibrium the CME is absent \cite{Valgushev:2015pjn,Buividovich:2015ara,Buividovich:2014dha,Buividovich:2013hza,Z2016_1,Z2016_2,nogo,nogo2,BLZ2021}. The CME is back out of equilibrium even very close to equilibrium \cite{BLZ2022}. It is also widely believed that the CME manifests itself in the steady state existing in the presence of parallel electric and magnetic fields \cite{Nielsen:1983rb}. It then may be detected through its contribution to negative magnetoresistance of  Dirac semimetals \cite{ZrTe5}.   At the same time the CSE exists as true equilibrium phenomenon \cite{Gorbar:2015wya}.

Using lattice regularization the CSE has been considered via analytical methods in \cite{KZ2017,SZ2020}. This regularization also allows to use the non - perturbative numerical methods  \cite{Buividovich:2013hza}. In \cite{Brandt2022AnomalousTP} the results of lattice simulations for the QCD with $N_f = 2+1$   at finite temperature have been presented. It appears that the CSE conductivity is suppressed above the crossover temperature. Its value increases with temperature. It increases very slowly, and as we will show, it approaches the conventional expression $1/(2\pi^2)$ per Dirac fermion at the Electroweak scale.

It is well - known that the theory with massless charged fermions is subject to dangerous infrared divergencies \cite{infrared1,infrared2,infrared3}. Interaction corrections to CSE have not been considered extensively so far. In \cite{Shovkovy} it has been argued that in QED the high orders of perturbation theory give corrections to the CSE conductivity for the system of massive fermions. However, the calculated corrections suffer from infrared divergencies, and depend on finite photon mass, which was introduced to avoid divergencies. To the best of our knowledge corrections to the CSE due to exchange by color gauge bosons have not been considered analytically.

The present paper is devoted to the consideration of interaction corrections to chiral separation effect. We assume, first of all, the corrections due to strong interactions in quark matter. However, the obtained  result on the topological expression for the CSE conductivity remains valid in any fermionic system with interactions, provided that the dangerous infrared divergences in the corrections to the CSE conductivity are absent. (The absence of such divergencies in QCD is provided by color magnetic confinement.) Our consideration is based on the lattice regularization of the QFT model. It remains valid, therefore, also for the consideration of the tight - binding models of solid state systems.

 As a useful tool we use Wigner - Weyl calculus in its form adopted for the lattice models. It allows to represent the conductivities of some non  - dissipative transport phenomena in the form of the topological invariants.
This formalism \cite{Weyl,Wigner} has been proposed originally as an alternative to conventional mathematical tools of  non-relativistic quantum mechanics \cite{Groenewold,Moyal}.  It has been extended also to be used in quantum field theory. The basic notions of this formalism are the Weyl symbol of operator and the Wigner distribution function. In Wigner - Weyl calculus the quantum state is described by Wigner distribution instead of a wave function. The product of operators is replaced by the so - called star (or,  Moyal) product of functions in phase space.

Wigner - Weyl calculus was proposed first for the continuous systems. The attempts to construct the analogoud formalism for the systems defined on discrete lattice faced certain difficulties \cite{Schwinger,Buot1,Buot2,Buot3,Wootters,Leonhardt,Kasperkovitz,Ligabo}.  A version of lattice Wigner - Weyl calculus has been proposed recently in \cite{FZ2019_2}. In this version basic properties of Weyl symbols of operator and Moyal product repeat precisely those of the continuous Wigner - Weyl calculus. In the present paper we rely on the simplified form of this calculus \cite{Zhang_Zubkov_PRD_2019,Suleymanov_Zubkov_2019,Fialk_Zubkov_2020_sym,Zhang_Zubkov_JETP_2019,Zhang_Zubkov_PhysLet_2020}. In this form the main properties of the Weyl symbols and Moyal product are not precise, and are approximate. This calculus may be used  if the inhomogeneity is sufficiently weak. In solid state physics the approximate Wigner - Weyl calculus may be used when magnetic field is much smaller than $10^5$ Tesla, i.e. in all realistic situations (maximal value of magnetic field accessed in present experiments does not exceed $100$ Tesla). In lattice regularized quantum field theory the approximate Wigner - Weyl calculus may be applied when the model approaches continuum limit.

We consider lattice models of rather general type with fermions placed in the four - component Dirac spinors. For quark matter these spinors carry extra internal indices - color and flavor. In condensed matter systems the extra indices have the meaning of valley, spin, etc. Action for the fermions contains the $4\times 4$ matrices to be expressed through Dirac matrices $\gamma^k$  ( $k=1,2,3,4,5$), and their derivatives $\sigma^{kj} = \frac{i}{4}[\gamma^k,\gamma^j]$. It will always be needed that at low energies the models under consideration obey chiral symmetry. This means that matrix $\gamma^5$ commutes or anti - commutes with the one - particle Hamiltonian in a small vicinity of the Fermi surfaces (Fermi points).  Fermi surface may be understood as the position of the singularities in momentum space for the two - point Green function $\hat{G} = \hat{Q}^{-1}$. We will call $\hat{Q}$ the lattice Dirac operator.
In the presence of weak inhomogeneity the notion of Fermi surface is to be replaced by the coordinate - dependent Fermi surface \cite{Volovik2003}. From the mathematical point of view inhomogeneity might be considered as weak if Wigner transformed  Green function $G_W(p,x)$ has singularities for any value of $x$ very close to the positions of the zeros of Weyl symbol $Q_W(p,x)$ of operator $\hat{Q}$. This will give the definition of the Fermi surface in case of weak inhomogeneity.

In \cite{SZ2020} it has been shown if low energy effective theory obeys chiral symmetry, the axial current of CSE in the non - homogeneous system of general type is still proportional to external magnetic field. Being averaged over the whole volume of the system its response to the chemical potential may be expressed as
\begin{equation}
	\frac{d}{d\mu}\bar{J}_5^k = \frac{\mathcal N}{4\pi^2}\epsilon^{ijk0}  F_{ij}\label{1_}
\end{equation}
where $\mathcal{N}$ is a topological invariant expressed through $G_W$ and $Q_W$.
\begin{widetext}
	\begin{eqnarray}
		\mathcal{N}
		&=&\frac{1}{48 \pi^2 {\bf V}}
		\int_{\Sigma_3}
		\int d^3x
		\tr \Bigg[\gamma^5
		{G}_W\star d { Q}_W \star { G}_W
		\wedge \star d { Q}_W\star { G}_W \star \wedge d { Q}_W
		\Bigg]\label{Ncompl0}
	\end{eqnarray}
\end{widetext}
$Q_W$ and $G_W$ here are defined for the system without external magnetic field. The above expression has been derived for the non - interacting system.
\mzo{ Surface $\Sigma_3$ in momentum space consists of the two hyperplanes $p_4 = \pm \epsilon \to 0$.  It  surrounds the singularities of an expression standing inside the integral.
 $\mathcal{N}$ is robust to smooth modification of the system if this modification does not break chiral symmetry around the positions of mentioned singularities \footnote{\mzo{ 
 Eq. (\ref{Ncompl0}) may be also rewritten in the form, when surface $\Sigma_3$ has other form (but still surrounds the singularities of expression standing inside the integral). Then the derivatives with respect to spatial coordinates $\vec{p}$ and $\vec{x}$ entering the Moyal (star product) should act on the functions depending also on $p_4$, but the latter is to be considered as function of $\vec{p}$, $\vec{x}$ determined by the form of $\Sigma_3$.  Besides, in this expression instead of $G_W(x,p)$ we should substitute the solution $\tilde{G}(\vec{x},\vec{p})$ of equation $Q_W(x,(\vec{p},p_4(\vec{p},\vec{x})))\star \tilde{G}(\vec{x},\vec{p})$. It coincides with the Wigner transformed Green function only in the simplest case when hypersurface $\Sigma_3$ corresponds to $p_4 = \pm \epsilon \to 0$. More details on the way how the star - product is to be understood in this case will be given below in Sect. \ref{topinv}, when we will consider the similar expression for the interacting systems.}}.}
 
 For the sufficiently weak homogeneity  the value of $\mathcal{N}$ may be calculated easily, and is given by the number of the species of chiral Dirac fermions in the low energy effective theory. As a result the CSE conductivity is given by ${N_c N_f}/(2\pi^2)$ for the system of $N_f$ non - interacting quarks ($N_c=3$ is the number of colors).

In the present paper we prove that Eqs. (\ref{1_}) and (\ref{Ncompl0}) remain valid in the interacting system, when the Green function $G$ is replaced by the  complete interacting renormalized Green function. Correspondingly, $\hat{Q}$ is defined as its inverse.  In QCD this may be applied directly to the region of phase diagram with small temperature and large values of $\mu$, if at those values of $\mu$ the chiral symmetry is restored, while gap is not opened, i.e. there is no color superconductivity. Besides, the same result may be applied to the electronic quasiparticles in Weyl semimetals, which simulate high energy physics in laboratory. Here we need $\mu \gg T$.

In QCD at finite temperature Eqs. (\ref{1_}) and (\ref{Ncompl0}) may be applied to the quark - gluon plasma phase provided that temperature may be neglected compared to the quark chemical potential. In advance it is not clear to which degree $\mu$ has to be larger than $T$ in order to apply the topological expression. In order to clarify this we calculate the non  - perturbative contributions to CSE conductivity using method of field correlators developed by Yu.A. Simonov and collaborators \cite{Simonov2002vva,Simonov2009zx,Simonov2007jb,Krivoruchenko2010jz,Abramchuk2019,OrlovskySimo,Agasian2017,Andreichikov2017ncy,Agasian2006ra,Simonov2016xaf}. At the  values of $\mu$ and $T$ around $T_c$ the $SU(3)$ coupling constant is $\alpha_s(T_c) \approx 0.3$, and the perturbative corrections are expected to contribute by amount of about $30$ per cent. We disregard these contributions completely. Therefore, we pretend here on the qualitative estimate only in the region $T \sim \mu \sim T_c$. However, at larger values of $T$ and/or  $\mu$ $\alpha_s$ decreases, at the energy scale of the order of $1$ GeV it becomes of the order of $0.2$, then the perturbative corrections become smaller. The obtained results demonstrate that in the phase diagram (see Fig. \ref{phase_diagram}) the topological result remains far from the region accessible by modern colliders. For example, in the region of phase diagram, where RHIC operates, the CSE conductivity is suppressed by the factor $\sim 2$ compared to the conventional value.

\section{Wigner - Weyl formalism in lattice theory in the presence of interactions. }

\subsection{Partition function of non - interacting system}

First we consider the non - interacting fermion system in lattice regularization. We refer here to the system of quarks. However, the obtained results do not depend on the nature of the fermions, and they remain valid for any fermionic system with chiral fermions.

In Euclidean space-time the partition function is expressed through the inverse bare Green function. It will be called further the Dirac operator and denoted by  $\hat Q$. The partition function  is given by
\be
Z = \int D\bar{\psi }D\psi
\,\, e^{S[\psi ,\bar\psi  ]}
\label{Z01}
\ee
Here $\psi, \bar{\psi}$ are the Grassmann - valued quark fields, while $S$ is the action
\begin{equation}\begin{aligned}
		&S[\psi ,\bar\psi  ]=\int_{\mathcal M} \frac{d^D{p}}{|\mathcal M|}\bar\psi({p}) \hat Q(i\partial_{p},{p})\psi({p})=\\
		&\int_{\mathcal M} \frac{d^D{p}}{|\mathcal M|}\bar\psi^a({p}) \hat Q^{ab}(i\partial_{p},{p})\psi^b({p})=\\
		&\sum_{{r}_n} \int_{\mathcal M} \frac{d^D{p}}{|\mathcal M|}  Q_W^{ab}({r}_n,{p}) W^{ba}({r}_n,{p})=\\
		&\sum_{{r}_n} \int_{\mathcal M} \frac{d^D{p}}{|\mathcal M|}  \tr  \Big[ Q_W({r}_n,{p})  W({r}_n,{p})\Big]
		\end{aligned}  \end{equation}
where we used Weyl symbols of operators
\begin{equation} \begin{aligned}
		{Q}^{}_W(x,p) \equiv \int_{\cal M} dq e^{ix q} \langle {p+q/2}| \hat{Q} | {p-q/2}\rangle\label{Q_W}
	\end{aligned}\,,
\end{equation}
and
\be W({r}_n,{p})=\(\ket{\psi}\bra{\psi}\)_W \ee
Here by $\ket{\psi}\bra{\psi}$ we denote operator with Grassmann - valued matrix elements
${\psi(x)}{\bar{\psi}}(y)$.
For simplicity of notations we discretize both space coordinates and imaginary time.  In the case of condensed matter system we are able to take off the discretization of imaginary time in order to arrive at the conventional expression $\hat{Q} = i \omega - \hat{H}$, where $\hat{H}$ is one - particle Hamiltonian.

Using  \textit{Peierls} substitution in the presence of slowly varying gauge field  (\ref{Q_W}) takes the form
\be Q_W(r,{p})\rightarrow Q_W(r,{p}-{A}(r))\label{QWA} \ee
Here direct dependence on $r$ is caused by the other slow varying external fields.
Partition function receives the form
\begin{equation} \begin{aligned}
		Z=\int D\bar\psi D\psi  \exp
		\left(-\sum_{{r}_n} \int_{\mathcal M|} \frac{d^D{p}}{|\mathcal M|}  \tr  \Big[ Q_W({r}_n,{p})  W({r}_n,{p})\Big]\right)
		\end{aligned} \end{equation}
Propagator of fermions is defined as
\begin{equation}
	\hat G=-\frac{1}{Z}\int D\bar\psi D\psi \ket{\psi} \bra{\bar\psi} \exp\left(\int \frac{d^D{p}}{|\mathcal M|}\bar\psi({p}) \hat Q(i\partial_{p},{p})\psi({p})\right)
	 \end{equation}
Its expression in momentum space is
\begin{equation}\begin{aligned}
		&G({p}_1,{p}_2)=\bra{{p}_1} G \ket{{p}_2}\\&=
		\frac{1}{Z}\int D\bar\psi D\psi \bar\psi({p}_2) \psi({p}_1) \exp\left(\int \frac{d^D{p}}{|\mathcal M|}\bar\psi({p}) \hat Q(i\partial_{p},{p})\psi({p})\right)
		\end{aligned}  \end{equation}
	
\subsection{Partition function for the system with interactions}	
	
Let us consider the case of  interactions between the fermions. As above we speak first of all of the systems, in which the fermions are quarks, while the gauge group is the color $SU(3)$ of strong interactions. However, the expressions to be derived further are valid also for the other interactions, and the other fermions with chiral symmetry.  This brings the partition function to the form	
\begin{equation} \begin{aligned}
		Z&=\int D\bar\psi D\psi DA e^{-S_A[A]} \\&\exp
		\left(-\sum_{{r}_n} \int_{\mathcal M|} \frac{d^D{p}}{|\mathcal M|}  \tr  \Big[ Q_W({r}_n,{p-A})  W({r}_n,{p})\Big]\right)
		\end{aligned} \end{equation}
Here $A$ is $SU(3)$ gauge field, while $S_A[A]$ is the pure gauge field action, \mzz{which contains the gauge fixing term (we are speaking here of the gauge group $SU(3)$ only).}

Variation of partition function may be expressed as follows
\begin{equation} \begin{aligned}
		&\delta \log Z=\\
		&-\frac{1}{Z}
		\int D\bar\psi D\psi DA e^{-S_A[A]}
		\Big[\sum_{{r}_n} \int \frac{d^D{q}}{|\mathcal M|}
		\delta Q^{ab}_W({r}_n,{q-A}) W^{ba}({r}_n,{q})\Big]
		\\&\exp \left(-\sum_{{r}_n} \int \frac{d^D{p}}{|\mathcal M|}  Q^{ab}_W({r}_n,{p-A})  W^{ba}({r}_n,{p})\right)=\\
		&- \sum_{{r}_n} \int DA e^{-S_A[A]} \frac{d^D{q}}{|\mathcal M|}
		\delta Q^{ab}_W({r}_n,{q-A})
		\nonumber\\&\Bigl[ \frac{1}{Z} \int D\bar\psi D\psi  W^{ba}({r}_n,{q-A})
		\\& \exp \left(-\sum_{{r}_n} \int \frac{d^D{p}}{|\mathcal M|}  Q^{ab}_W({r}_n,{p-A})  W^{ba}({r}_n,{p})\right) \Bigr]=\\
		& \sum_{{r}_n} \int DA e^{-S_A[A]} \frac{d^D{q}}{|\mathcal M|}
		\delta Q^{ab}_W({r}_n,{q-A}) G^{ba}_W({r}_n,{q-A})=\\
		& \sum_{{r}_n} \int DA e^{-S_A[A]} \frac{d^D{q}}{|\mathcal M|}
		\tr \left[ \delta Q_W({r}_n,{q-A}) G_W({r}_n,{q-A}) \right]
		\end{aligned} \end{equation}
In the case when dependence of $\delta Q_W({r}_n,{q-A})$ on $A$ may be neglected, we obtain
\begin{equation} \begin{aligned}
		\delta \log Z&= \sum_{{r}_n} \int \frac{d^D{p}}{|\mathcal M|}
		\tr \left[ \delta Q_W({r}_n,{p}) {\bf G}_W({r}_n,{p}) \right]\\
		&= \int d^D x  \int \frac{d^D{p}}{{\bf v}|\mathcal M|}
		\tr \left[ \delta Q_W(x,{p}) {\bf G}_W(x,{p}) \right]\\
		&= \int d^D x  \int \frac{d^D{p}}{(2\pi)^D}
		\tr \left[ \delta Q_W(x,{p}) {\bf G}_W(x,{p}) \right]
		\label{dlogZ}\end{aligned} \end{equation}
Here $\bf v$ is the elementary lattice cell volume. In the second line assume that expression standing under the sum depends slowly on $r_n$. In the last line we use that ${\bf v} |{\mathcal M}| = (2\pi)^D$. We rewrite this expression as
\begin{equation} \begin{aligned}
		&\delta \log Z= \tr \left[\hat {\bf G} \delta \hat Q  \right]=\Tr[{\bf G}_W\star \delta Q_W]=
		\Tr[ {\bf G}_W\delta Q_W]
		\end{aligned} \end{equation}
	with
	$$
	\star = e^{\frac{i}{2} \left( \overleftarrow{\partial}_{x}\overrightarrow{\partial_p}-\overleftarrow{\partial_p}\overrightarrow{\partial}_{x}\right )}
	$$
Here {\bf by $\bf G$ we denote the complete interacting two - point quark Green function} while $G$ is the Green function in the presence of external $SU(3)$ field $A$. \mzz{Notice that $\bf G$ is the quark Green function calculated in certain gauge of the $SU(3)$ group. Our further results do not depend on the particular choice of the gauge.} The star may be removed here if $\delta Q_W(p,x)$ as a function of $x$ is localized in finite region of space.  In the following we will always denote by bold letters the complete Green function and its inverse, while ordinary letters will denote bare quantities with no interactions taken into account.\\
%{\red
%We can use the analytical continuation ${r}_n\rightarrow x$ in order to perform the derivatives in the $\star$-product (ref Precise WW - Zubkov+Fialkovsky). We will denote the coordinates on the lattice as $x\in {\mathcal O}$.
%\be
%{\mathcal O}=\{an,n\in \mathbb{Z}\}
%\ee
%where $a$ is the lattice length.
%}
{
	From now on we use continuum limit for the coordinates ${r}_n\rightarrow x$. This is possible if variations of fields on the distances of the order of lattice spacings are neglected. }\\
In the presence of an extra external gauge field we substitute $p\rightarrow p-{\cal A}$
\be
Q_W(x,p-A)\rightarrow Q_W(x,p-A-{\cal A})
\ee
Variation with respect to the external gauge field ${\cal A}\rightarrow {\cal A}+\delta {\cal A}$ gives
\begin{eqnarray}
&& Q_W(x,p-A-({\cal A}+\delta {\cal A}))=Q_W(x,p-A-{\cal A})\nonumber\\&&+\pd_{{\cal A}_i}Q_W(x,p-A-{\cal A})\delta {\cal A}_i
\end{eqnarray}
and
\be
\delta Q_W=\pd_{{\cal A}_i} Q_W \delta {\cal A}_i=-\pd_{p_i} Q_W \delta {\cal A}_i
\ee
Expression for $\partial_{p_i} Q_W(x,p-A)$ obviously becomes independent of $ A$, when we approach the continuum limit. Since we are interested in continuum limit of lattice theory, the electric current may be taken in the form
\be
j_i(x)=\frac{\delta \log Z}{\delta {\cal A}_k(x)}=
-\int_{(2\pi)^D} \frac{d^Dp}{|\mathcal M|}
\tr \left[ {\bf G}_W(x,p) \partial_{p_i} Q_W(x,p)  \right]
\label{j_i}
\ee
Based on analogy with electric current the naive expression for local axial current density may be defined as
\be
j^5_k(x)= -
\int_{\mathcal M} \frac{d^Dp}{(2\pi)^D}
\tr \left[ \gamma^5  {\bf G}_W(x,p) \pd_{p_k} Q_W(x,p)  \right]
\ee

\subsection{Gauge transformation of Weyl symbol}

\label{AWS}

$U(1)$ gauge transformation acts as
$\ket{x} \to e^{i \alpha(x)} \ket{x}$. As a result Weyl symbol of an operator $\hat{B}$ is transformed as
\begin{eqnarray}
	B_W(x,p) &=& \int dy e^{-i y p} \langle {x+y/2}| \hat{B} | {x-y/2}\rangle\nonumber\\ & \to &
	\int dy e^{-i y p + i \alpha(x+y/2) - i \alpha(x-y/2)}\nonumber\\&& \langle {x+y/2}| \hat{B} | {x-y/2}\rangle
\end{eqnarray}
Here we replace the sum over lattice points by an integral because we assume that all fields vary slowly, so that their variation at the distance of lattice spacing may be neglected. Let us consider those gauge transformations, for  which function $\alpha$ almost does not vary at the distances of the order of the correlation length $\lambda$ characterizing operator $\hat{B}$, i.e. $|\lambda \partial \alpha|\ll 1$. We call these transformations "slow" (with respect to $\hat B$). For them we obtain:
\begin{eqnarray}
	B_W(x,p)  & \to &
	\int dy e^{-i y p + i \alpha(x+y/2) - i \alpha(x-y/2)}\nonumber\\&& \langle {x+y/2}| \hat{B} | {x-y/2}\rangle\nonumber\\ & \approx & \int dy e^{-i y (p -  \partial \alpha(x))}\nonumber\\&& \langle {x+y/2}| \hat{B} | {x-y/2}\rangle\nonumber\\&=&	B_W(x,p - \partial \alpha(x))
\end{eqnarray}
If operator $\hat B$ depends on the $U(1)$ gauge field $A$ then we may require that the gauge transformation of $\hat B$ should be compensated by the gauge transformation of field $A$. This occurs, for example, for Dirac operator $\hat{Q}$ due to gauge invariance of the whole model. Consideration of "slow" gauge transformation results in the requirement that Weyl symbol $B_W(x,p)$ depends on $A(x)$ through the functional dependence on $p - A(x)$, and gauge invariant quantities: field strength $F_{ij}$ and its derivatives, provided that variation of $A(x)$ may be neglected at the distances of the order of $\lambda$, i.e. $|\lambda^2 F_{ij}| \ll 1$. As a result for such $A(x)$ we may represent $B_W$ as a series
\begin{eqnarray}
	B_W(x,p)&=&B^{(0)}_W(x,p - A(x))+B^{(1)}_{(ij) W}(x,p - A(x))F_{ij}(x)\nonumber\\&&+B^{(2)}_{(ijk) W}(x,p - A(x))\partial_kF_{ij}(x) + ...\label{BWA}
\end{eqnarray}
Here dots denote the higher  order terms in derivatives.
This expansion is reasonable, i.e. the higher order terms are smaller than the lower order terms under the same condition $|\lambda^2 F_{ij}| \ll 1$.

 In particular, for bare $\hat Q$ the correlation length $\lambda$ is given by the lattice spacing, and we arrive at Eq. (\ref{QWA}) for the fields $A$ that vary slowly at the distance of the order of lattice spacing.

\subsection{Renormalized quark velocity and renormalized axial current}
\label{SectRenorm}
{The meaning of $-\partial_{p_i} Q_W(x,p-A)$ is matrix of bare quark velocity. It is natural that the electric current is (up to electric charge of quark) given by averaging of quark velocity.  A natural supposition is that in quantum theory the renormalized quark velocity has to be substituted to this expression. Namely, let us denote by $\bf Q$ an operator inverse to $\bf G$, which is the complete quark Green function with interactions taken into account. \mzz{Notice again that both $\bf G$ and $\bf Q$ are to be calculated after the gauge fixing procedure for $SU(3)$ gauge group.} Then
the renormalized velocity operator is
\begin{equation}
v_R=-\partial_{p_i} {\bf Q}_W(x,p-A)
\end{equation}	
It can be shown using the methodology developed in \cite{ZZ2022} that to all orders in perturbation theory the  electric current averaged over the system volume $V$ is given by
$$
\frac{1}{\beta V} \int d^D x j_k(x) =
-\frac{1}{\beta V} \int_{\mathcal M} d^D x \frac{d^Dp}{(2\pi)^D}$$$$
\tr \left[ {\bf G}_W(x,p) \partial_{p_i} {\bf Q}_W(x,p)  \right]
$$
The latter expression does not have much sense because according to the Bloch theorem the persistent current vanishes in non - marginal systems. The proof of the theorem follows from the fact that the above expression is a topological invariant.
At the same time the above expression does not mean that the local current density may be expressed through the renormalized velocity operator
$$
 j_k(x) \ne
- \int_{\mathcal M}  \frac{d^Dp}{(2\pi)^D}
\tr \left[ {\bf G}_W(x,p) \partial_{p_i} {\bf Q}_W(x,p)  \right]
$$
The non - trivial expression appears, however, when we consider response of the above expression to external fields. In this way considering the electric field that has equal values but opposite directions in the two pieces of space, in \cite{ZZ2022} it has been shown that the (integer) Hall conductivity (averaged over the system area) does not have perturbative corrections, and may be expressed through the complete interacting Green functions. As a result, we can take the "renormalized" expression for the electric current density
$$
{\bf j}_k(x) =
- \int_{\mathcal M}  \frac{d^Dp}{(2\pi)^D}
\tr \left[ {\bf G}_W(x,p) \partial_{p_i} {\bf Q}_W(x,p)  \right]
$$
and calculate its response to electric field. This way the correct expression for the Hall conductance is reproduced, while the longitudinal contribution vanishes. We conclude that for the calculation of the physical observables averaged over the whole system area in  the QHE systems the renormalized expression for the current density may be used.

\mz{Based on an analogy to electric conductivity below we accept as the definition of the renormalized axial current the expression with the  operator of renormalized velocity in place of the bare velocity:
\be
{\bf j}^5_k(x)= -
\int_{\mathcal M} \frac{d^Dp}{(2\pi)^D}
\tr \left[ \gamma^5  {\bf G}_W(x,p) \pd_{p_k} {\bf Q}_W(x,p)  \right]\label{J5R}
\ee
For the calculation of the electric current for the QHE system integral over the whole Brillouin zone is to be calculated. We will see that contrary to this for the calculation of CSE conductivity one should integrate in momentum space along the infinitely small hypersurface surrounding the position of Fermi surface/Fermi point.  As a result in the field theory with spatial isotropy at zero temperature and zero chemical potential (for example, in QED) we need expression for $\pd_{p_k}{\bf Q}_W(x,p)$ in the small vicinity of $p = 0$.  There we have
$$
\pd_{p_k}{\bf Q}_W(x,p) \approx \gamma^k Z_F
$$
Here $Z_F$ is the fermion field renormalization constant. One can see, therefore, that at $T=\mu=0$ the only difference if we substitute to the CSE current the renormalized velocity (instead of the bare one) is appearance of the renormalization constant $Z_F$. However, this is precisely what is to be done for the calculation of the renormalized axial current at $T=\mu=0$ (if the latter is defined as $\langle \bar{\Psi}_R \gamma^\mu \gamma^5 \Psi_R\rangle$, where $\Psi_R = Z_F^{1/2} \Psi$ is the renormalized field operator, while $\Psi$ is bare fermionic field). Notice, that unlike vector current the axial current is not Noether current responsible for the transport of a conserved charge. Therefore, its definition in the interacting systems is flexible.  In the present paper we extend definition of Eq. (\ref{J5R}) to the systems with nonzero $T$ and $\mu$, and, to the systems with spatial anisotropy. As it was mentioned above, Eq. (\ref{J5R}) will be considered below as the {\it definition of renormalized axial current}.  }

\subsection{Groenewold equation and its iterative solution}
(Renormalized) Dirac operator and (renormalized) Green function obey the following equation
\be \hat {\bf Q} \hat {\bf G}=1 \ee
Weyl-Wigner transformation results in the Groenewold equation
\be {\bf Q}_W(p,x) \star  {\bf G}_W(p,x)=1 \label{Groen}\ee

We assume here that all external fields vary slowly, i.e. these variations may be neglected at the distance of the order of lattice spacing. Then Weyl symbol of bare (non - interacting) Dirac operator has the functional dependence  ${ Q}_W(p-A(x),x)$ in the presence of external field $A_i(x)$ (corresponding to the field strength $F_{ij}$). Here the coordinate dependence caused by the other external fields is given by direct dependence on $x$.
Function ${\bf Q}_W(p,x)$ with interaction corrections can be represented as
\begin{equation}
	{\bf Q}_W(p,x) = {\bf Q}^{(0)}_W(p-A(x),x) + {\bf Q}^{(1)}_{(ij) \,W}(p-A(x),x)F_{ij} + ...\label{QpA}
\end{equation}
Dots represent the terms proportional to the higher powers of $F$ and the derivatives of $F$. As it was explained above in Sect. \ref{AWS}, this expansion is valid under the condition $|\lambda^2 F_{ij}| \ll 1$, where $\lambda$ is the correlation length associated with the given interacting system. This expansion is reasonable, at least, when we consider the DC CSE conductivity, i.e. the response of the axial current to sufficiently small external magnetic field. Recall that the correlation length associated with bare Dirac operator is equal to the lattice spacing. In the presence of interactions the correlation length may become much larger, of the order of the existing dimensional parameters of the system. For example, for the quark matter at zero temperature such parameters are the quark chemical potential and $\Lambda_{QCD}$. Assuming $\mu > \Lambda_{QCD}$, for these systems Eq. (\ref{QpA}) may be applied for magnetic fields much smaller than $\Lambda_{QCD}^2$.

In order to illustrate representation of Eq. (\ref{QpA}) let us consider  approximation, when only the one - gluon exchange is taken into account:
\begin{eqnarray}
	&&{\bf Q}^{(0)}_W(p-A(x),x) \approx  {Q}_W(p-A(x),x) \\&& - g^2\int \frac{d^Dk}{(2\pi)^D} {\cal D}^{(0)ab}_{\mu\nu}(k) \, \gamma^\mu t_a G^{(0)}_W(p-k-A(x),x)\gamma^\nu t_b \nonumber\\
	&&{\bf Q}^{(1)}_{(ij)\,W}(p-A(x),x)\approx  -g^2\int \frac{d^Dk}{(2\pi)^D}\nonumber\\&& {\cal D}^{(0)ab}_{\mu\nu}(k) \,\gamma^\mu t_a G^{(1)}_{(ij)\,W}(p-k-A(x),x)\gamma^\nu t_b\nonumber
\end{eqnarray}
Here ${\cal D}^{(0)}$ is gluon propagator, while $t_a$ are the Gell - Mann matrices, $a,b$ are color indexes. Above we denote by $G^{(0)}_W(p,x)$ solution of reduced Groenewold equation (i.e. the one without $A(x)$):
$$
G^{(0)}_W(p,x)\star Q_W(p,x)=1
$$
At the same time the first order term in derivative of $A$ is
$$
{G}_{(ij)W}^{(1)}=
\frac{i}{2} \Big[ {G}_W^{(0)}\star \(\pd_{p_i} {Q}_W^{(0)}\) \star {G}_W^{(0)}
\star \(\pd_{p_j} {Q}_W^{(0)}\) \star {G}_W^{(0)} \Big]
$$
It gives the first order term (expansion over derivatives of $A$) in solution of equation
$$
(G^{(0)}_W(p,x)+{G}_{(ij)W}^{(1)}F_{ij})\star Q_W(p-A(x),x)=1
$$
This result follows the derivation presented in \cite{ZW2019} and is based on expansion
$$
\star = 1+\frac{i}{2} \left( \overleftarrow{\partial}_{x}\overrightarrow{\partial_p}-\overleftarrow{\partial_p}\overrightarrow{\partial}_{x}\right )+...
$$
Among the second order diagrams let us consider the representative one, in which the gluon propagator receives correction from the quark loop. This gives the following contribution to quark self energy up to the terms linear in $F_{ij}$:
\begin{eqnarray}
&&	\Delta\Sigma = g^2\int \frac{d^Dk}{(2\pi)^D} {\cal D}^{(1)ab}_{\mu\nu W}(k,x) \, \gamma^\mu t_a G^{(0)}_W(p-k-A(x),x)\gamma^\nu t_b \nonumber\\
	&&+g^2\int \frac{d^Dk}{(2\pi)^D}F_{ij} {\cal D}^{(0)ab}_{\mu\nu W}(k,x) \,\gamma^\mu t_a G^{(1)}_{(ij)\,W}(p-k-A(x),x)\gamma^\nu t_b\nonumber
\end{eqnarray}
Here $-\Delta\Sigma$ is one of the many terms entering perturbative expansion of ${\bf Q}_W(p,x)$. The corresponding contribution to gluon propagator is
\begin{widetext}
\begin{eqnarray}
	 {\cal D}^{(1)ab}_{\mu\nu W}(k,x) &=& g^2 {\cal D}^{(0)ac}_{\mu\rho W}(k) \star \int \frac{d^D q}{(2\pi)^D} {\rm tr} \Big(G_W(-q,x)\gamma^\rho t^b G_W(k-q,x) \gamma^\sigma t^d \Big)\star {\cal D}^{(0)db}_{\sigma\nu W}(k)\nonumber\\
	 &=& g^2{\cal D}^{(0)ac}_{\mu\rho W}(k) \star \int \frac{d^D q}{(2\pi)^D} {\rm tr} \Big(G^{(0)}_W(-q+A(x),x)\gamma^\rho t^b G^{(0)}_W(k-q-A(x),x) \gamma^\sigma t^d \Big)\star {\cal D}^{(0)db}_{\sigma\nu W}(k) \nonumber\\ && + g^2{\cal D}^{(0)ac}_{\mu\rho W}(k) \star \int \frac{d^D q}{(2\pi)^D} {\rm tr} \Big(G^{(0)}_W(-q+A(x),x)\gamma^\rho t^b G^{(1)}_{(ij)W}(k-q-A(x),x) \gamma^\sigma t^d \Big)\star {\cal D}^{(0)db}_{\sigma\nu W}(k)F_{ij}
	 \nonumber\\ && - g^2{\cal D}^{(0)ac}_{\mu\rho W}(k) \star \int \frac{d^D q}{(2\pi)^D} {\rm tr} \Big(G^{(1)}_{(ij)W}(-q+A(x),x)\gamma^\rho t^b G^{(0)}_{W}(k-q-A(x),x) \gamma^\sigma t^d \Big)\star {\cal D}^{(0)db}_{\sigma\nu W}(k)F_{ij}\nonumber\\
	 &=& g^2{\cal D}^{(0)ac}_{\mu\rho W}(k) \star \int \frac{d^D q}{(2\pi)^D} {\rm tr} \Big(G^{(0)}_W(-q,x)\gamma^\rho t^b G^{(0)}_W(k-q,x) \gamma^\sigma t^d \Big)\star {\cal D}^{(0)db}_{\sigma\nu W}(k) \nonumber\\ && + g^2{\cal D}^{(0)ac}_{\mu\rho W}(k) \star \int \frac{d^D q}{(2\pi)^D} {\rm tr} \Big(G^{(0)}_W(-q,x)\gamma^\rho t^b G^{(1)}_{(ij)W}(k-q,x) \gamma^\sigma t^d \Big)\star {\cal D}^{(0)db}_{\sigma\nu W}(k)F_{ij}
	 \nonumber\\ && - g^2 {\cal D}^{(0)ac}_{\mu\rho W}(k) \star \int \frac{d^D q}{(2\pi)^D} {\rm tr} \Big(G^{(1)}_{(ij)W}(-q,x)\gamma^\rho t^b G^{(0)}_{W}(k-q,x) \gamma^\sigma t^d \Big)\star {\cal D}^{(0)db}_{\sigma\nu W}(k)F_{ij}
\end{eqnarray}
\end{widetext}
In the last three rows we performed the shift of variable $q -A(x) \to q$.  One can see that still there are the two types of the contributions to ${\bf Q}_W(p,x)$ described by Eq. (\ref{QpA}). Obviously the same consideration may be extended to all orders of perturbation theory. The same refers also to the non - perturbative contributions to  ${\bf Q}_W(p,x)$ according to the arguments presented in Sect \ref{AWS}.

In the similar way the solution of Groenewold equation (\ref{Groen}) for the interacting Green function (up to the terms linear in $F$) is given by
$$
{\bf G}_W(p,x) \approx {\bf G}^{(0)}_W(p,x)+{\bf G}_{(ij)W}^{(1)}F_{ij}
$$
where  ${\bf G}^{(0)}_W(p,x)$ is solution of reduced Groenewold equation (i.e. the one without $A(x)$):
$$
{\bf G}^{(0)}_W(p,x)\star {\bf Q}^{(0)}_W(p,x)=1
$$
The first order term in derivative of $A$ is more complicated than in case of non - interacting Green function:
\begin{eqnarray}
{\bf G}_{(ij)W}^{(1)}&=&
\frac{i}{2} \Big[ {\bf G}_W^{(0)}\star \(\pd_{p_i} {\bf Q}_W^{(0)}\) \star {\bf G}_W^{(0)}
\star \(\pd_{p_j} {\bf Q}_W^{(0)}\) \star {\bf G}_W^{(0)} \Big]\nonumber\\ && - {\bf G}_W^{(0)}\star {\bf Q}_{(ij)W}^{(1)} \star {\bf G}_W^{(0)} 
\end{eqnarray}

\section{Topological expression for chiral separation effect in the presence of interactions at $T=0$}

\subsection{Response of axial current to magnetic field}

As has been explained above, the local (renormalized) axial current density is given by
\be
{\bf j}^5_k(x)= -
\int_{\mathcal M} \frac{d^Dp}{(2\pi)^D}
\tr \left[ \gamma^5 {\bf G}_W(x,p) \pd_{p_k} {\bf Q}_W(x,p)  \right]
\label{ji5x}\ee
We obtain the following term with the linear response to external field strength:
\be
{\bf j}_k^5(x)&=-
\frac{i}{2}
\int_{\mathcal M} \frac{d^Dp}{(2\pi)^D}
\tr \Bigl[\gamma^5
\Big[ {\bf G}_W^{(0)}\star \(\pd_{p_i} {\bf Q}_W^{(0)}\) \\&\star {\bf G}_W^{(0)}
\star \(\pd_{p_j} {\bf Q}_W^{(0)}\) \star {\bf G}_W^{(0)} \Big] \pd_{p_k}{\bf Q}_W^{(0)}
\Bigr]
F_{ij}\\
& + \int_{\mathcal M} \frac{d^Dp}{(2\pi)^D}
\tr \Bigl[\gamma^5
\Big[ {\bf G}_W^{(0)}\star {\bf Q}_{(ij)W}^{(1)} \star {\bf G}_W^{(0)}  \Big] \pd_{p_k}{\bf Q}_W^{(0)}
\Bigr]
F_{ij}
\\
& - \int_{\mathcal M} \frac{d^Dp}{(2\pi)^D}
\tr \Bigl[\gamma^5 {\bf G}_W^{(0)}   \pd_{p_k} \Big[{\bf Q}_{(ij)W}^{(1)}\Big]
\Bigr]
F_{ij}
\label{ji5lr}\ee
Averaging the local current over the whole system volume we get
\be
&\bar{J}_i^5\equiv \frac{1}{\beta {\bf V}}\sum_x {\bf j}_i^5(x)\\&=
-\frac{1}{\beta {\bf V}}\int d^Dx \int_{\mathcal M} \frac{d^Dp}{{\bf v}|\mathcal M|}
\tr \left[\gamma^5  {\bf G}_W(x,p) \partial_{p_i} {\bf Q}_W(x,p)  \right]\\
&=-\frac{1}{\beta {\bf V}}\Tr \left[ \gamma^5 {\bf G}_W(x,p) \partial_{p_i} {\bf Q}_W(x,p)  \right]
\label{Ii5}\ee
Here $\bf v$ is volume of the lattice cell. We have a useful formula ${\bf v}|\mathcal M| = (2\pi)^D$.
We obtain
\be
\bar{J}_k^5&=
-\frac{i}{2}\frac{1}{\beta{\bf V}}
\int d^Dx\int_{\mathcal M} \frac{d^Dp}{(2\pi)^D}\tr \Bigl[\gamma^5
\Bigl[ {\bf G}_W^{(0)}\star \(\pd_{p_i} {\bf Q}_W^{(0)}\) \\&\star {\bf G}_W^{(0)}
\star \(\pd_{p_j} {\bf Q}_W^{(0)}\) \star {\bf G}_W^{(0)}\pd_{p_k}{\bf Q}_W^{(0)} \\&
 + 2i {\bf G}_W^{(0)}\star {\bf Q}_{(ij)W}^{(1)} \star {\bf G}_W^{(0)} \pd_{p_k}{\bf Q}_W^{(0)}
\\& - 2i  {\bf G}_W^{(0)}   \pd_{p_k} \Big[{\bf Q}_{(ij)W}^{(1)}\Big] \Bigr]
\Bigr]
F_{ij}
\label{Ji5lr}\ee

%\newpage

\subsection{Axial current for massless fermions at finite temperature}

For the sake of regularization and also because we are going to consider QCD at finite temperature, we introduce finite temperature.
Matsubara frequencies are
$
p_4=\omega_n=\frac{2\pi\(n+\frac{1}{2}\)}{\beta}
$.
Here the inverse temperature $\beta = 1/T$ is taken in lattice units:
$
N_t\equiv\frac{1}{T}
$,
and the values of $p_4$ are
$
p_4=\frac{2\pi\(n_4+\frac{1}{2}\)}{N_t}$, $ n_4=-\frac{N_t}{2},..,\frac{N_t}{2}-1
$.
The boundary values are
$
\om_{n=-\frac{N_t}{2}}=\frac{2\pi\(-\frac{N_t}{2}+\frac{1}{2}\)}{N_t}=
-\pi+\frac{\pi}{N_t}
$ and
$
\om_{n=\frac{N_t}{2}-1}=\frac{2\pi\(\frac{N_t}{2}-\frac{1}{2}\)}{N_t}=
\pi-\frac{\pi}{N_t}
$.
The Matsubara frequencies most close to zero are:
$
\om_{n=0}=\frac{\pi}{N_t}
$ and
$
\om_{n=-1}=-\frac{\pi}{N_t}
$.
One can see that $\om_n$ never equals to zero. Therefore, the propagator does not have poles in momentum space. The axial current receives the form
\be
&\bar{J}_k^5=-
\frac{i}{2}\frac{1}{\beta{\bf V}}
\sum_{n=-\frac{N_t}{2}}^{\frac{N_t}{2}-1}
\int d^3x \int_{\mathcal M_3} \frac{d^3p}{(2\pi)^3}\\
&\tr \Bigl[\gamma^5
\Bigl[ {\bf G}_W^{(0)}\star \(\pd_{p_i} {\bf Q}_W^{(0)}\) \\&\star {\bf G}_W^{(0)}
\star \(\pd_{p_j} {\bf Q}_W^{(0)}\) \star {\bf G}_W^{(0)}\pd_{p_k}{\bf Q}_W^{(0)} \\&
+ 2i {\bf G}_W^{(0)}\star {\bf Q}_{(ij)W}^{(1)} \star {\bf G}_W^{(0)} \pd_{p_k}{\bf Q}_W^{(0)}
\\& - 2i  {\bf G}_W^{(0)}   \pd_{p_k} \Big[{\bf Q}_{(ij)W}^{(1)}\Big] \Bigr]
\Bigr]
F_{ij}
\label{Ji5}\ee

%\subsection{Variation with respect to chemical potential}
Chemical potential may be introduced as $\omega_n\ra\omega_n-i\mu$. Therefore, the response of axial current to variation of chemical potential $\delta \mu$ and to external field strength $F_{\mu \nu}$ receives the form
\be
&\bar{J}_k^5=-
\frac{1}{{2\bf V}\beta}
\sum_{n=-\frac{N_t}{2}}^{\frac{N_t}{2}-1}
\int d^3x \int_{\mathcal M_3} \frac{d^3p}{(2\pi)^3}\\
&\pd_{\omega_n}\tr \Bigl[\gamma^5
\Bigl[ {\bf G}_W^{(0)}\star \(\pd_{p_i} {\bf Q}_W^{(0)}\) \\&\star {\bf G}_W^{(0)}
\star \(\pd_{p_j} {\bf Q}_W^{(0)}\) \star {\bf G}_W^{(0)}\pd_{p_k}{\bf Q}_W^{(0)} \\&
+ 2i {\bf G}_W^{(0)}\star {\bf Q}_{(ij)W}^{(1)} \star {\bf G}_W^{(0)} \pd_{p_k}{\bf Q}_W^{(0)}
\\& - 2i  {\bf G}_W^{(0)}   \pd_{p_k} \Big[{\bf Q}_{(ij)W}^{(1)}\Big] \Bigr]
\Bigr]
F_{ij}\delta\mu
\label{Ji5mu1}\ee
 We represent the above expression as
\be
\bar{J}_k^5(x)=\mathcal{\sigma}_{ijk}F_{ij}\delta\mu
\label{Ji5mu2}\ee
where
\be
&\mathcal{\sigma}_{ijk}=
-\frac{1}{{2\bf V}\beta}
\sum_{n=-\frac{N_t}{2}}^{\frac{N_t}{2}-1}
\int d^3x \int_{\mathcal M_3} \frac{d^3p}{(2\pi)^3}\\
&\pd_{\omega_n}\tr \Bigl[\gamma^5
\Bigl[ {\bf G}_W^{(0)}\star \(\pd_{p_i} {\bf Q}_W^{(0)}\) \\&\star {\bf G}_W^{(0)}
\star \(\pd_{p_j} {\bf Q}_W^{(0)}\) \star {\bf G}_W^{(0)}\pd_{p_k}{\bf Q}_W^{(0)} \\&
+ 2i {\bf G}_W^{(0)}\star {\bf Q}_{(ij)W}^{(1)} \star {\bf G}_W^{(0)} \pd_{p_k}{\bf Q}_W^{(0)}
\\& - 2i  {\bf G}_W^{(0)}   \pd_{p_k} \Big[{\bf Q}_{(ij)W}^{(1)}\Big] \Bigr]
\Bigr]
\label{Nijk5}\ee
has the meaning of the CSE conductivity when external field strength corresponds to a constant magnetic field $H$: $F_{ij} = -\epsilon_{ijk} H_k$. Then
$$
\bar{J}_k^5(x)=-\mathcal{\sigma}_{ijk}\epsilon_{ijk^\prime} H_{k^\prime}\delta\mu
$$
We represent expression for the CSE conductivity as
\be
\mathcal{\sigma}_{ijk}=&
\sum_{n=-\frac{N_t}{2}}^{\frac{N_t}{2}-1}
\pd_{\om_n}\mathcal{\sigma}_{ijk}^{(3)}
\label{Nijk5_1}\ee
where
\be
\mathcal{\sigma}_{ijk}^{(3)}&=-
\frac{1}{{2\bf V}}
\int d^3x \int_{\mathcal M_3} \frac{d^3p}{(2\pi)^3}\\
&\tr \Bigl[\gamma^5
\Bigl[ {\bf G}_W^{(0)}\star \(\pd_{p_i} {\bf Q}_W^{(0)}\) \\&\star {\bf G}_W^{(0)}
\star \(\pd_{p_j} {\bf Q}_W^{(0)}\) \star {\bf G}_W^{(0)}\pd_{p_k}{\bf Q}_W^{(0)} \\&
+ 2i {\bf G}_W^{(0)}\star {\bf Q}_{(ij)W}^{(1)} \star {\bf G}_W^{(0)} \pd_{p_k}{\bf Q}_W^{(0)}
\\& - 2i  {\bf G}_W^{(0)}   \pd_{p_k} \Big[{\bf Q}_{(ij)W}^{(1)}\Big] \Bigr]
\Bigr]
\label{N_3}\ee

\subsection{The limit of small temperature }

The limit of small temperature $T\ra 0$, $N_t\ra \infty$, $\frac{\pi}{N_t}=\ep\ra 0$ allows to replace the sum by an integral. Value $\omega = 0$ is to be excluded from this integral:
\be
\sum_{n=-\frac{N_t}{2}}^{\frac{N_t}{2}-1} \tab\ra \tab
\frac{\beta}{2\pi}\int_{-\pi+\ep}^{0-\ep}d\omega+ \frac{\beta}{2\pi}\int_{0+\ep}^{\pi-\ep}d\omega
\ee
Then (\ref{Nijk5}) becomes
\be
\mathcal{\sigma}_{ijk}&=
\lim_{\ep\ra0}
\int_{-\pi+\ep}^{0-\ep}d\om
\pd_{\om}\mathcal{\sigma}_{ijk}^{(3)}
+
\int_{0+\ep}^{\pi-\ep} d\om
\pd_{\om}\mathcal{\sigma}_{ijk}^{(3)}\\
&=\lim_{\ep\ra0}\Big[
\mathcal{\sigma}_{ijk}^{(3)}(-\pi+\ep) - \mathcal{\sigma}_{ijk}^{(3)}(0-\ep)\\&+
\mathcal{\sigma}_{ijk}^{(3)}(0+\ep) - \mathcal{\sigma}_{ijk}^{(3)}(\pi-\ep)\Big]
\label{Nijk_int}\ee
using that $
\mathcal{\sigma}_{ijk}^{(3)}(-\pi)=\mathcal{\sigma}_{ijk}^{(3)}(\pi)
$, we obtain
\be
&\mathcal{\sigma}_{ijk}=
\lim_{\ep\ra0}\[
\mathcal{\sigma}_{ijk}^{(3)}(0+\ep) +\(-\mathcal{\sigma}_{ijk}^{(3)}(0-\ep)\) \]
\label{Nijk_N_3}\\
\ee
where
\begin{widetext}
	\be
	{\sigma}_{ijk}^{(3)}(\om=0\pm \ep)&=-\frac{1}{{2\bf V}}
	\int d^3x \int_{\mathcal M_3} \frac{d^3p}{(2\pi)^4}
	\tr \Bigl[\gamma^5
	\Bigl[ {\bf G}_W^{(0)}\star \(\pd_{p_i} {\bf Q}_W^{(0)}\) \star {\bf G}_W^{(0)}
	\star \(\pd_{p_j} {\bf Q}_W^{(0)}\) \star {\bf G}_W^{(0)}\star \pd_{p_k}{\bf Q}_W^{(0)} \\&
	+ 2i {\bf G}_W^{(0)}\star {\bf Q}_{(ij)W}^{(1)} \star {\bf G}_W^{(0)} \star \pd_{p_k}{\bf Q}_W^{(0)}
	 - 2i  {\bf G}_W^{(0)} \star  \pd_{p_k} \Big[{\bf Q}_{(ij)W}^{(1)}\Big] \Bigr]
	\Bigr]\Bigg|_{\om=0\pm \ep}
	\label{N_3_1}\ee
\end{widetext}

We are considering equilibrium theory, when both $\bf G$ and $\bf Q$ do not depend on time. As a result the singularities are situated at $\omega = 0$. The integrals avoid these singularities due to finite $\epsilon$. In the absence of inhomogeneity (when the stars may be omitted in the above expressions) at $\omega = 0$ the singularities of expressions standing in the integrals mark positions of Fermi surfaces.

\subsection{CSE conductivity as a topological invariant}

\label{topinv}

In Eq. (\ref{Nijk_N_3}) inside the integrals the two surfaces $\omega = \pm \epsilon$ cancel each other except for the small vicinity of the singularities. Therefore, we restrict integration in Eq. (\ref{N_3_1}) by the small regions (in the Brillouin zone) around the singularities.  The important assumption here is the presence of precise chiral symmetry in these regions. In the other words, the continuum limit of the lattice theory under consideration is chiral invariant.

Thus  $\gamma^5$ commutes/anti - commutes with $\bf Q$ and $\bf G$ inside the above expression for the CSE conductivity.
As a result the last two terms in Eq. (\ref{N_3_1}) cancel each other, while the sum of the integrals in Eq. (\ref{Nijk_N_3}) represents a topological invariant. It does not depend on the form of the surface in $4D$ momentum space surrounding the singularities. We deform this surface in such a way that it becomes small and surrounds the singularities. Therefore, instead of the two  infinitely close planes we may integrate over the sphere in momentum space (see the figure).

Thus we obtain
\begin{widetext}
	\be
	&\mathcal{\sigma}_{ijk}=\\
	&-\frac{1}{{2\bf V}}
	\int_{\Sigma_3} \frac{d^3p}{(2\pi)^4}
	\int d^3x
	\tr \Bigg[\gamma^5
	\Big[ {\bf G}_W^{(0)}\star \(\pd_{p_{[i}} {\bf Q}_W^{(0)}\) \star {\bf G}_W^{(0)}
	\star \(\pd_{p_{j]}} {\bf Q}_W^{(0)}\) \star {\bf G}_W^{(0)} \Big] \pd_{p_k}{\bf Q}_W^{(0)}
	\Bigg]
	\ee
\end{widetext}
Here the integral is over $\Sigma_3$, which is the 3D hypersurface in 4D momentum space that consists of the two infinitely close pieces of the planes.  $\gamma^5$ commutes/anti - commutes with $\bf G$ and $\bf Q$ in this region, and we rewrite this expression as
$$\sigma_{ijk} = -\epsilon_{ijk} \sigma_{CSE}/2
$$
with
\begin{equation}
	\sigma_{CSE} = \frac{\mathcal{N}}{2\pi^2}\label{sigmaH}
\end{equation}
and
\begin{widetext}
	\begin{eqnarray}
		\mathcal{N}&=&\frac{\epsilon_{ijk}}{48 \pi^2 {\bf V}}
		\int_{\Sigma_3} {d^3p}
		\int d^3x
		\tr \Bigg[\gamma^5
		\Big[ {\bf G}_W^{(0)}\star \(\pd_{p_{i}} {\bf Q}_W^{(0)}\) \star {\bf G}_W^{(0)}
		\star \(\pd_{p_{j}} {\bf Q}_W^{(0)}\) \star {\bf G}_W^{(0)} \Big] \pd_{p_k}{\bf Q}_W^{(0)}
		\Bigg]=\nonumber\\
		&=&\frac{1}{48 \pi^2 {\bf V}}
		\int_{\Sigma_3}
		\int d^3x
		\tr \Bigg[\gamma^5
		{\bf G}_W^{(0)}\star d {\bf Q}_W^{(0)} \star {\bf G}_W^{(0)}
		\wedge \star d {\bf Q}_W^{(0)}\star {\bf G}_W^{(0)} \star \wedge d {\bf Q}_W^{(0)}
		\Bigg]\label{eq47}
	\end{eqnarray}
\end{widetext}
This expression is topological invariant provided that $\gamma^5$ commutes or anti - commutes with ${\bf Q}_W$ and ${\bf G}_W$ in the vicinity of  \mzo{the singularities of the expression standing in the integral. Recall that here $\Sigma_3$ consists of the two hyperplanes $p_4 = \pm \epsilon$ (where $\epsilon \to 0$)} \footnote{\mzo{For the more general form of $\Sigma_3$  certain clarifications should be added.  
		More specifically, in Eq. (\ref{eq47}) we should understand function ${\bf Q}^{(0)}_W$ as depending on spatial momenta $\vec{p}$ and spatial coordinates $\vec{x}$, i.e. it should be substituted by  
		\begin{align}
			{\bf Q}^{(0)}_W\to\tilde{\bf Q}^{(0)}(\vec{x},\vec{p}) & \equiv {\bf Q}^{(0)}_W((\vec{x},0),(\vec{p},\pm\omega_{\pm}(\vec{p},\vec{x})) ) \label{eq1}\tag{A}
		\end{align}
		Here the three - dimensional vector $\vec{p}$ parametrizes surface $\Sigma_3(x)$. Function $\omega_\pm$ represents the dependence of Matsubara frequency on $\vec{p}$: $p_4 = \pm \omega_\pm(\vec{p},\vec{x})$, the upper sign is to be chosen for the upper piece of $\Sigma_3$, the lower sign - for the lower piece.  We took into account that in the considered equilibrium systems $\bf Q$ as well as $\bf G$ does not depend on $x^4$.}
		
		\mzo{In turn,  ${\bf G}^{(0)}_W$ in Eq. (\ref{eq47}) should be substituted by the $\star$ - inverse with respect to $\tilde{\bf Q}^{(0)}$, i.e.  ${\bf G}^{(0)}_W \to \tilde{\bf G}^{(0)}$ that obeys
		\begin{align}
			\tilde{\bf Q}^{(0)}(\vec{x},\vec{p}) e^{\frac{i}{2}(\overleftarrow{\partial}_{\vec{x}}\overrightarrow{\partial}_{\vec{p}}-\overleftarrow{\partial}_{\vec{p}}\overrightarrow{\partial}_{\vec{x}})}\tilde{\bf G}^{(0)}(\vec{x},\vec{p})=1\label{eq2} \tag{B}
		\end{align}
		With these clarifications Eq. (\ref{eq47}) reads (for a general form of $\Sigma_3$ depending on $x$): 
		%\begin{widetext}
			\begin{align}
				\mathcal{N}
				=&\frac{\epsilon_{ijk}}{48 \pi^2 {\bf V} }
				\int   {d^3 x} \int_{\Sigma_3(x)}d^3 p
				\tr \Bigl[\gamma^5
				\tilde{\bf G}^{(0)}\star \partial_{p_i} \tilde{\bf Q}^{(0)}\nonumber\\& \star\tilde{\bf G}^{(0)}
				\star \partial_{p_j} \tilde{\bf Q}^{(0)}\star \tilde{\bf G}^{(0)} \star  \partial_{p_k} \tilde{\bf Q}^{(0)}
				\Bigr]\label{eq3}\tag{C}
			\end{align}
		%\end{widetext}
		with the three - dimensional volume ${\bf V} = \int d^3 x$, which is assumed to be large. With this clarification Eq. (\ref{eq47}) represents topological invariants robust to both smooth modification of the system and smooth modification of the form of hypersurface $\Sigma_3$. In both cases the singularities should be avoided, i.e. modifying the system one should not pass over a phase transition, while modifying $\Sigma_3$ one should not cross the position of  singularities that extends the notion of Fermi surface to the interacting non - homogeneous systems. 
 }}. \mzo{Recall that the superscript $^{(0)}$ means that we set $A = 0 $ inside ${\bf Q}_W$ and ${\bf G}_W$.}

In the particular case, when background is homogeneous, we obtain:
\begin{widetext}
	\begin{eqnarray}
		\mathcal{N}&=&\frac{1}{48 \pi^2 }
		\int_{\Sigma_3}
				\tr \Bigg[\gamma^5
		{\bf G}_W^{(0)} d {\bf Q}_W^{(0)}  {\bf G}_W^{(0)}
		\wedge  d {\bf Q}_W^{(0)} {\bf G}_W^{(0)}  \wedge d {\bf Q}_W^{(0)}
		\Bigg]
	\end{eqnarray}
\end{widetext}
Here  $\bf Q = {\bf G}^{-1}$. Index $^{(0)}$ means that magnetic field and chemical potential are set to zero to calculate the Green function. For the most simple case of the Fermi point, when chemical potential is zero, the form of $\Sigma_3$ here is an infinitely small three - dimensional sphere surrounding $p=0$.

At zero temperature and large baryonic chemical potential the quark - gluon system may enter the quarkyonic phase with restored chiral symmetry. The increase of chemical potential may also lead to formation of color superconductivity. In the hypothetical phase, where the chiral symmetry is restored, while the color superconductivity is not yet formed, the above mentioned topological invariant $\cal N$ counts the number of chiral Dirac fermions
$$
{\sigma}_{CSE} = \frac{N_c N_f}{2 \pi^2}
$$
where $N_c = 3$, while $N_f$ is the number of quarks with masses smaller than $\mu$.
We then come to the standard expression for the CSE conductivity in this phase. This result, presumably, may be valid for matter existing within the neutron stars.

The above results also allow to predict the same expression for one Dirac fermion $\sigma_{CSE} = \frac{1}{2 \pi^2}$ in the Weyl semimetal at zero temperature. These materials realize within solid state physics the systems of relativistic fermions. Here electronic quasiparticles are in place of quarks, while Coulomb interactions substitute the exchange by  $SU(3)$ gauge bosons. Those Coulomb interactions may be strong due to the electric permittivity. Besides, in these materials there are the other interactions between electrons. The result on the topological expression for the CSE conductivity does not depend on the nature of inter - fermions interactions. Therefore, the given topological expression remains valid in these systems as well.

\section{Non - perturbative corrections to CSE conductivity in QCD at finite temperature}

In this section we confirm the predictions of the previous sections by direct calculations using method of field correlators developed by Yu.A. Simonov and collaborators (see also \cite{A2023}).

\subsection{Representation of axial current through the sum over quark trajectories, $T> T_c$}

We start from the following expression for the \mz{bare} axial current at the temperature above the deconfinement crossover
\begin{eqnarray}
     &&\braket{j^5_\mu(x)} =  \braket{\tr_{c,D} \gamma_5\gamma_\mu S^{(\text{reg})}(x,x)}\\ &
     = & \frac{\braket{{\rm Det}\,(\slashed D(B, \cA) + m)\, \tr_{c,D} \gamma_5\gamma_\mu (\slashed D(B, \cA) + m)^{-1}_{xx}}_B}{\braket{{\rm Det}\,(\slashed D(B, \cA) + m)}_B}\nonumber
\end{eqnarray}
\mz{Since the renormalization of the axial current operator occurs at small distances, it is the perturbative phenomenon. We neglect it here completely as we are interested in the non - perturbative contributions to the CSE conductivity originated from large distances.} 
 
For brevity we restrict ourselves in this section by contribution of one quark flavor. The loop is also regularized by finite temperature.
${\rm tr}_c$ stands for the trace over color and Dirac spinor indices, respectively.
$\braket{\ldots}_B$ stands for averaging over thermodynamic ensemble
with temperature $T>T_c, \,(\beta=T^{-1})$ above the deconfinement crossover temperature $T_c\sim 160$ MeV,
at non-zero baryon density that is defined with the given quark flavor chemical potential $\mu$,
in the gluonic background field $B$
($B_\mu=B_\mu^a t_a$, $t_a$ are the generators of SU(3) algebra in fundamental representation),
in external constant magnetic field
$\nabla\times\vec\cA = \vec H$ directed along axis Z. Then $H_3 = \cF_{12} = \text{const}$
is the only non-zero component of the field strength tensor.
Chemical potential is introduced as imaginary part of the fourth component of electromagnetic potential (in Euclidean space) $iq\cA_4 = \mu$,
where $q$ is electric charge of the given quark flavor.

In the following we apply quenched approximation, in which the fermion determinant is dropped. For the calculation of quark propagator we use world-line formalism:
\begin{align}
    \braket{j^5_\mu(x)} &\approx \langle \tr_{c,D} \gamma_5\gamma_\mu (-\slashed D(B, \cA) + m)_{x} \nonumber\\&
        \int_0^{+\infty}ds~\xi(s)~(\overline{\cD^4z})_{xx}^s e^{-m^2s-K} \times \\
        &\times P_FP_B\exp\large(ig\oint B\cdot dz + iq\oint\cA\cdot dz
        \nonumber\\&+ \int_0^sd\tau\sigma^{\rho\sigma}(gF_{\rho\sigma}(z,z_0) + q\cF_{\rho\sigma})\large)\rangle_B,\nonumber
\end{align}
where $\sigma_{\rho\sigma} = \frac{i}{4}[\gamma_\rho,\gamma_\sigma]$ is the generator of $SO(3,1)$.
The covariant derivative is $D(B, \cA) = \partial - igB - iq\cA$. Here $g$ is the $SU(3)$ coupling constant.
Function $\xi(s)$ regularizes the loop integral. It is needed to remove singularity at $s=0$. We also denote here
$$
K =  \int_0^{s} d\tau \Bigl(\frac{\dot{z}^2(\tau)}{4} + m^2\Bigr)
$$

The path integral $(\overline{\cD^4z})_{xy}^s$ describes quark motion from point $x$ to $y$ in world-line (proper) time $s$.
The anti-periodic boundary conditions are assumed for the fermion
when the trajectory wraps around the (imaginary) time direction. Euclidean space is taken in the form $\mathbb{R}^3\times S^1$, where the direction of imaginary time is a circle $S^1$ of length $\beta=T^{-1})$.
The path integral discretization is implied here \cite{OrlovskySimo}:
\begin{align}
    &(\overline{\cD^4z})_{xy}^s = \lim_{N\to +\infty}\prod_{m=1}^{N}\frac{d^4 z_m}{(4\pi\varepsilon)^2}
    \sum_{n=-\infty}^{+\infty}(-1)^n \frac{d^4p}{(2\pi)^4}\nonumber\\&
    e^{(ip_\mu(\sum_{i=1}^{m}\D z^\mu_i - (x-y) - n\beta\delta^\mu_4))}, \, \varepsilon=s/N.
    \label{EqPathInt}
\end{align}

Trace over the Dirac indices is simplified due to the presence of $\gamma_5$. We consider the leading order in external field and use that
$\tr_D\gamma_5\gamma_\mu\gamma_\nu\gamma_\lambda\gamma_\rho = - 4\epsilon_{\mu\nu\lambda\rho}$
(in Euclidean space-time  $\{\gamma_\mu,\gamma_\nu\} = 2\delta_{\mu\nu}$).
The leading order in magnetic field is given by
\begin{widetext}
\begin{align}
    \braket{j^5_\mu(x)} \approx &
     \langle \tr_{c,D} \gamma_5\gamma_\mu\gamma_\lambda\frac{i}{4}[\gamma_\rho,\gamma_\sigma] ~(-D_\lambda(B, \cA))_x
        \int_0^{+\infty}ds~\xi(s)~(\overline{\cD^4z})_{xx}^s e^{-m^2s-K} \times \nonumber\\
        &\times P_B\exp\left(ig\oint B\cdot dz + iq\oint\cA\cdot  dz\right)
        ~ \int_0^sd\tau(gF_{\rho\sigma}(z,z_0) + q\cF_{\rho\sigma})\rangle_B
\end{align}
\end{widetext}
We disregard here influence of external magnetic field on configurations of gluonic fields $B$ \cite{Abramchuk2019}. As a result
the gluonic field correlators  $\braket{\ldots}_B$ are isotropic. Besides, we disregard the spin - gluon interactions (these interactions do not contribute much, for example, to the string tension).
Due to the definite direction of external magnetic field the Dirac indices are  $(\mu\lambda[\rho\sigma]=34[12])$.
The proper time integral $\int_0^s d\tau \cF_{12} = s~H$ represents insertion of  electromagnetic vertex at the points along  quark trajectory
\begin{widetext}
\begin{align}
    \braket{j^5_\mu(x)} \approx
    4i\delta_{\mu 3}q H \int_0^{+\infty}\xi(s)sds~\langle&\tr_c(D_4(B, \cA))_x(\overline{\cD^4z})_{xx}^s
        e^{-m^2s-K}\times \nonumber\\
        &\times P_B\exp\left(ig\oint B\cdot dz + iq\oint\cA\cdot  dz\right)\rangle_B.\label{jH}
\end{align}
\end{widetext}
The covariant derivative acts on path integral. Definition of parallel transporter
results in
\begin{align}
    &(D_\lambda(B,\cA))_x P_B\exp\left(ig\int^x_y B\cdot dz + iq\int_y^x\cA\cdot  dz\right) = 0,\nonumber\\& \quad z(0)=x.
\end{align}
As a result the covariant derivative is reduced to the ordinary one that acts on $(\overline{\cD^4z})_{xx}^s$. The major temperature dependent contributions to the integral over the trajectories come from the loops that wrap $n$ times around $S^1$, and we replace
 $\partial_4\to\frac{\partial}{\partial(n\beta)}$. The contribution to the axial current not dependent on temperature may be neglected. At zero temperature the chiral symmetry breaking suppresses the chiral separation effect.

The electromagnetic part standing in exponent of the Wilson loop (for the quark trajectory $z^{(n)}$ that wraps $n$ times around the temporal direction) is
\begin{align}
    \quad q\oint\cA\cdot dz^{(n)} = q\Phi_H -i\mu n \beta.
\end{align}
Here $\Phi_H$ is the magnetic flux through the spatial projection of the loop.
We are considering the linear response to magnetic field, Eq. (\ref{jH}) is proportional to $H$, and we disregard the external magnetic field in the remaining expression. We come to the following expression
\begin{widetext}
\begin{align}
	\braket{j^5_\mu(x)} &\approx
	4i\delta_{\mu 3}q H \int_0^{+\infty}\xi(s)sds~\prod_{m=1}^{N}\frac{d^4 z_m}{(4\pi\varepsilon)^2}
	\sum_{n=-\infty}^{+\infty}(-1)^n \frac{d^4p}{(2\pi)^4}\frac{\partial}{\partial(n\beta)}\Big(
	e^{(ip_\mu(\sum_{i=1}^{m}\D z^\mu_i - (x-y) - n\beta\delta^\mu_4))}\nonumber\\ &
	e^{-m^2s-K}\Big)\times \langle\tr_c P_B\exp\left(ig\oint B\cdot dz + \mu n \beta \right)\rangle_B\nonumber\\ &=
	4i\delta_{\mu 3}q H \int_0^{+\infty}\xi(s)sds~\prod_{m=1}^{N}\frac{d^4 z_m}{(4\pi\varepsilon)^2}
	\sum_{n=-\infty}^{+\infty}(-1)^n \frac{d^4p}{(2\pi)^4}\frac{\partial}{\partial(n\beta)}\Big(
	e^{(ip_\mu(\sum_{i=1}^{m}\D z^\mu_i - (x-y) - n\beta\delta^\mu_4))}\nonumber\\ &
	e^{-m^2s-K}\Big)\times \langle\tr_c P_B\exp\left(ig\oint B\cdot dz + \mu n \beta \right)\rangle_B.\label{jH2}
\end{align}
\end{widetext}
The last expression allows to represent the CSE conductivity $\frac{\partial}{\partial \mu}\frac{\partial}{\partial H_k}\braket{j^5_k(x)}$ as
 \begin{align}
 	\sigma_{CSE} &\approx
 	4 i \int_0^{+\infty}\xi(s)sds~\prod_{m=1}^{N}\frac{d^4\D z_m}{(4\pi\varepsilon)^2}
 	\sum_{n=-\infty}^{+\infty}(-1)^n \frac{d^4p}{(2\pi)^4}\nonumber\\& \langle\tr_c P_B\exp\left(ig\oint B\cdot dz  \right)\rangle_B e^{ \mu n \beta}\nonumber\\&
 	\frac{\partial}{\partial({\rm log}\,\beta)} e^{-m^2s-K+(ip_\mu(\sum_{i=1}^{m}\D z^\mu_i - (x-y) - n\beta\delta^\mu_4))} .\label{jH3}
 \end{align}

\subsection{Evaluation of integral over quark trajectories}

Below we adopt version of Simonov technique proposed in \cite{Z2020}. It is based on the Abelian Diakonov - Petrov representation of Wilson loop. In this representation we obtain the following expression for quark condensate (which is a function of bare mass $m$, temperature, and chemical potential)
\begin{widetext}
  \begin{align}
 	\sigma_{CSE} &\approx
 	-4 i N_c  \frac{\partial}{\partial m^2} 	 \int_0^{+\infty}\xi(s)ds~\prod_{m=1}^{N}\frac{d^4\D z_m}{(4\pi\varepsilon)^2}
 	\sum_{n=-\infty}^{+\infty}(-1)^n \frac{d^4p}{(2\pi)^4}e^{ \mu n \beta} \nonumber\\ & \frac{\partial}{\partial({\rm log}\,\beta)}\Big(
 	e^{-m^2s-K+(ip_\mu(\sum_{i=1}^{m}\D z^\mu_i - (x-y) - n\beta\delta^\mu_4))}
 	\Big)\times \langle  \exp\left(ig\oint {\cal B}\cdot dz  \right)\rangle_B.\label{jH4}
 \end{align}
\end{widetext}
Here kernel $K$ is redefined:
$$
K =  \int_0^{s} d\tau \Bigl(\frac{\dot{z}^2(\tau)}{4} + m^2\Bigr) - \kappa \int_0^{s} \sqrt{\dot{z}^2(\tau)}d\tau
$$
Constant $\kappa$ entering this expression contains ultraviolet divergency, and is to be absorbed by renormalization of quark mass.  Abelian field $\cal B$ is defined as the component of the $SU(3)$ gauge field (taken in fundamental representation):
$$
{\cal B}_\mu = B^{11}_\mu
$$
In this abelian representation the Wilson loop $\cal C$ factorizes in the simply - connected loop ${\cal C}_l$ (the one, which does not wrap along the $S^1$), and the straight Polyakov line ${\cal L}^{(n)}$ that  is wrapped $n$-times along the $S^1$
\begin{align}
    \braket{W[\cal C]}_B = \braket{ W[{\cal C}_l]W[{\cal L}^{(n)}] }_B
\end{align}
Here
$$
W[{\cal C}] = \exp\left(ig \oint_{\cal C} {\cal B}\cdot dz  \right)
$$
Following \cite{Agasian2017} we neglect correlation between the Polyakov line and the remaining part of the Wilson loop. This results in
\begin{align}
	\braket{W[\cal C]}_B \approx \braket{ W[{\cal C}_l]}_B L^{(n)},
	\,L^{(n)} \approx L^{|n|}.
\end{align}
and
$$
L^{(n)} = \exp\left(ig\oint_{{\cal L}^{(n)}} {\cal B}\cdot dz  \right)
$$

The Polyakov line determines potential $V_1$
\begin{align}
    &L = \exp\left(-\frac{V_1(r\to +\infty, T)}{2T}\right),\nonumber\\&
    \, V_1(r\to +\infty, T) = V_1(T) = V_1
\end{align}
$V_1$ is  energy required to overcome the remnant interaction that bounds the quark as a part of a color-singlet state.
This potential was not yet calculated within the method of field correlators, and we use here the lattice data  \cite{Simonov2007jb}
\begin{align}
   & V_1(T>T_c) = \frac{175\text{ MeV}}{1.35 ~T/T_c - 1}, \nonumber\\&
     V_1(T_c) = 0.5\text{ GeV},
    \quad T_c = 160\text{ MeV}.\label{EqV1fit}
\end{align}

For the sake of rough evaluation we substitute $W[{\cal C}_l]$ by its spatial projection with the dominant contribution given by color - magnetic confinement $W[{\cal C}_l] \sim \exp(-\sigma_H S_3[\vec z])$,
where $S_3$ is the minimal area spanned on the spatial projection of Wilson loop. Effectively the color magnetic confinement results in the appearance of the thermal quark mass
\begin{align}
     &\red{m\to} \sqrt{m^2 + m_D^2/4}, \label{EqMD}
    \quad m_D^2 = c_D^2\sigma_H(T),
    \nonumber\\&\sigma_H(T)\approx c_\sigma^2g^4(T,\mu)T^2,
\end{align}
\red{%where $M$ is the resulting quark mass that contains both current mass $m$ and thermal mass $m_D/2$.
    Here $c_D\approx 2$ and $c_\sigma\approx 0.56$ are numerical constants that are taken from the analysis of the experimental consequences of the thermal mass appearance. Those values are extracted from lattice data \cite{Agasian2006ra}. For $T \approx \mu \approx T_c$ we have $m_D/2 \approx 320$ MeV.
    }

\red{
However,  non-perturbative ``perimeter-law'' contribution to quark self-energy shifts down the effective quark mass.
Let us adapt the result of \cite{Simonov2001iv} for the quark propagator to the deconfined phase
\begin{align}
    &\Delta m^2 = -\Lambda 
    = -\int d^4(y-x) \times\nonumber\\
    &\times\braket{\sigma_{\mu\nu} F^{\mu\nu}(x)\Phi_{xy}\sigma_{\mu'\nu'} F^{\mu'\nu'}(y)\Phi_{yx}}_B G(x,y).
    \label{EqNPqse}
\end{align}
In the deconfined phase the Color-Electric confining correlator is absent while the Color-magnetic is present
(in fact, we neglect all the correlators $D_1^{E,H,EH}$ but $D^H$)
$
\braket{\sigma_{\mu\nu} F^{\mu\nu}(x)\Phi_{xy}\sigma_{\mu'\nu'} F^{\mu'\nu'}(y)\Phi_{yx}} 
    \approx \braket{\sigma_{ij} F^{ij}(x)\Phi_{xy}\sigma_{i'j'} F^{i'j'}(y)\Phi_{yx}}.
$
}

\red{
Since the QCD vacuum correlation length $\lambda\sim 1\text{ GeV}^{-1} \ll \beta$ 
(in the old paper \cite{Simonov2001iv} the length is denoted as $T_g$)
in the interesting to us temperature range,
the integral for the non-perturbative self energy converges within one winding.
Also, the current quark masses for the light flavors are small in comparison to the inverse correlation length.
Thus we approximate the exact squared propagator  in the external magnetic field $G$ in \eqref{EqNPqse} with the free scalar propagator.
}

\red{
The consideration of \cite{Simonov2001iv} is applicable
up to the overall spin-averaging factor:
the factor $\sigma_{\mu\nu}\sigma^{\mu\nu}=D(D-1)/4$ in the confined phase 
is to be replaced with $\sigma_{ij}\sigma^{ij}=(D-1)(D-2)/4$.
Thus, the quark mass shift $\Delta m_q^2$ in the QGP phase is twice smaller then in the hadronic phase.
Finally the resulting effective quark mass $M$ is 
\begin{gather}
    \Delta m^2 \approx -\frac2\pi\sigma_{H}(T),\\
    \quad M^2 = m^2 + (c_D^2/4-2/\pi)\sigma_{H}(T). \label{EqMDnpQSE}
\end{gather}
The correction reduces the screened quark mass \eqref{EqMD} by a factor} \mzz{$\sqrt{1-\frac2\pi}\sim\frac23$. In particular, we then have $m_D/2 \approx 200$ MeV for $T \approx \mu \approx T_c$.
}
%%%%%%%%%%%%%%%%%%%%%%

Effect of the appearance of thermal mass may be taken into account roughly if the integration over the spatial coordinates of the quark trajectories is performed as for the free particle, but with the current mass $m$ substituted by $M$ (see \cite{Agasian2006ra,Agasian2017,Andreichikov2017ncy}). We set $K_3 = \int_0^s d\tau \dot{ \vec z}^2/4$, and obtain:
\begin{gather}
	\int({\cD^3\vec z})_{\vec x, \vec x}^s
	e^{-K_3-m^2s} \langle \exp\left(ig\oint \vec {\cal B}\cdot d\vec z \right)\rangle_B
	\sim \frac{e^{-M^2s}}{(4\pi s)^{3/2}}, \label{spat}
\end{gather}
The running coupling may be evaluated in one loop as
\begin{align}
	g^{2}(T,\mu) = \frac{1}{2b_0\log\frac{\sqrt{T^2 + 3 \mu^2/\pi^2}}{T_cL_\sigma}},
	\quad (4\pi)^2b_0 = \frac{11}{3}N_c - \frac23 N_f.
	\label{EqGT}
\end{align}
with  $L_\sigma \approx 0.1$. Numerical estimate for temperature and chemical potential around $T_c$ is
$$
\alpha_s(T_c,T_c) \approx 0.29
$$
This demonstrates that the perturbative corrections may, in principle, change the result by about 30 percent.

The calculation of the integral over the  temporal part of the quark trajectories takes into account the nonzero value of the Polyakov line. Here we denote $K_4 = \int_0^s d\tau \dot z_4^2/4$:
\begin{align}
	\int({\cD z_4})_{0,n\beta}^s
	e^{-K_4} \langle \exp \left(ig\int {\cal B}_4 dz_4 + iq\int\cA_4  dz_4\right)\rangle_B
	\approx\nonumber&\\
	\approx \frac{e^{-\frac{n^2\beta^2}{4s}}}{\sqrt{4\pi s}}L^{|n|}\exp(\mu n\beta )&.\label{temp}
\end{align}
In Appendix \ref{App1} we represent for comparison the calculation of path integral  for the free fermions.

Following \cite{Agasian2017} we conclude that here the result for the free fermions is to be used, where we substitute instead of the current mass of quark its thermal (Debye) mass and the Polyakov line. We combine Eqs. (\ref{spat}) and (\ref{temp}) to calculate the CSE conductivity:
\begin{widetext}
\begin{align}
%    = 4\delta_{\mu 3}N_cqH \int_0^{+\infty}sds~\xi(s)~\red{(i\partial_4)_x}(\overline{\cD^4z})_{xx}^s e^{-m^2s-K}
%        \braket{N_c^{-1}\tr_cW[z]}_B e^{\mu\int dz_4(\tau)}\\
    \sigma_{CSE} \approx
   - \frac{\partial}{\partial M^2} \frac{ N_c }{2\pi^2}  ~\int_0^{+\infty}\frac{ds}{s^2}
        \sum_{n=1}^{+\infty}(-1)^{n}\cosh(\mu n\beta )L^n \frac{\partial}{\partial({\rm log}\,\beta)}
        ~\exp\left(-M^2s - \frac{n^2\beta^2}{4s}\right)
\end{align}
\end{widetext}
In this expression only the nonperturbative contributions are taken into account.  The sub - dominant perturbative contributions are neglected here.
The non-winding trajectories, $n=0$, are not taken into account since this divergent contribution is $T$- and $\mu$-independent ``vacuum density'' \cite{Vilenkin}. Its derivative does not give contributions to $\sigma_{CSE}$.
Now the regularization $\xi(s)$ is no longer needed.\\

\begin{figure}[h]
	\centering  %
	\includegraphics[width=0.9\linewidth]{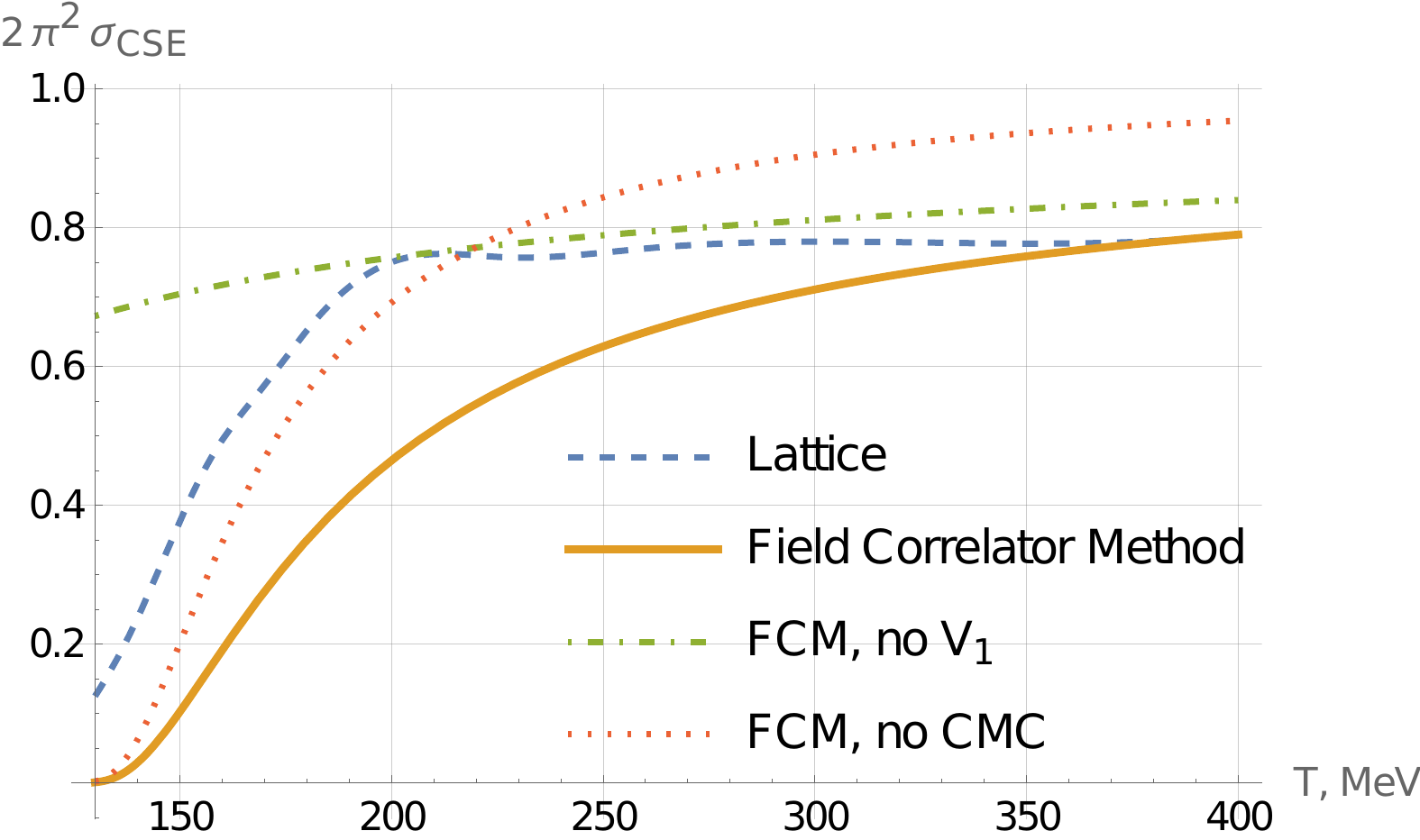}  %
    \caption{\red{We represent here the comparison of our non - perturbative calculation \eqref{EqICSE0} using method of field correlators with the lattice numerical simulations  taken from  \cite{Brandt2022AnomalousTP} calculated on the lattices $24^3 \times 6$ and $24^3 \times 8$. 
    The plot represents the data on $2 \pi \sigma_{CSE}$ 
    (per Dirac fermion, i.e divided by the number of quark flavors $N_f$ and colors $N_c$) 
    at $\mu = 0$.} \mzz{ The dashed line represents lattice data. Solid line represents the results obtained via field correlator method. Besides, we represent here the results obtained using two modifications of the FCM (field correlator method): the dotted line represents results with the thermal quark mass disregarded, i.e. without color magnetic confinement (CMC), while the dashed - dotted line represents the results with the Polyakov line contribution $V_1$ disregarded.  } 
    }  %
	\label{comparison}   %
\end{figure}

\subsection{Evaluation of $\sigma_{CSE}$}
We use integral representations for the modified Bessel functions:
\begin{gather}
    K_\nu(z) = \frac12\left(\frac{z}{2}\right)^\nu\int_0^{+\infty}
        \exp\left(-t-\frac{z^2}{4t}\right)\frac{dt}{t^{\nu+1}},\\
    K_0'(z) = -K_1(z),\\
    K_\nu(z) = \frac{\sqrt\pi(z/2)^\nu}{\Gamma(\nu+\frac12)}
        \int_0^{+\infty} e^{-z\cosh t}(\sinh t)^{2\nu}dt,
\end{gather}
To use the first expression we substitute  $s = M^2 t$:
\begin{align}
	\sigma_{CSE} &\approx -  \frac{\partial}{\partial M^2}
	\frac{2 M^2 N_c }{\pi^2} ~
	\sum_{n=1}^{+\infty}(-1)^{n}{\cosh(\mu n\beta )L^n }\nonumber\\&\frac{\partial}{\partial {\rm log}\, \beta}\frac{K_1(n \beta M)}{n \beta M}
\end{align}
Then we use the second representation
\begin{align}
	\sigma_{CSE} &\approx  -\frac{\partial}{\partial M^2}
	\frac{2 M^2 N_c }{\pi^2} ~
	\sum_{n=1}^{+\infty}(-1)^{n}{\cosh(\mu n\beta )L^n }\nonumber\\&\frac{\partial}{\partial {\rm log}\, \beta}\int_0^\infty e^{-n \beta M {\rm cosh}\, t}{\rm sinh}^2\,t\,  dt
\end{align}
and substitute here $p = M\sinh t$. As a result
\begin{widetext}
\begin{align}
   \sigma_{CSE} &\approx  \frac{\partial}{\partial M^2}
    \frac{N_c}{\pi^2} ~\sum_{n=1}^{+\infty}(-1)^{n}~
        \int_0^{+\infty}{p^2dp}
        \left(e^{\beta n(\mu - V_1/2 - \sqrt{p^2 + M^2})}\beta n  + (\mu\to-\mu)\right)\nonumber\\
       &=  -
       \frac{N_c}{2\pi^2} ~\sum_{n=1}^{+\infty}(-1)^{n}~
       \int_0^{+\infty}\frac{p^2dp}{\sqrt{p^2 + M^2}}
       \left(e^{\beta n(\mu - V_1/2 - \sqrt{p^2 + M^2})}(\beta n)^2  + (\mu\to-\mu)\right)
       \nonumber\\
       &=  -
       \frac{\partial^2}{\partial \mu^2}\frac{N_c}{2\pi^2} ~\sum_{n=1}^{+\infty}(-1)^{n}~
       \int_0^{+\infty}\frac{p^2dp}{\sqrt{p^2 + M^2}}
       \left(e^{\beta n(\mu - V_1/2 - \sqrt{p^2 + M^2})}  + (\mu\to-\mu)\right)
\end{align}
\end{widetext}
The sum is calculated as geometrical progression,
which yields  Fermi-Dirac distribution
$f_\beta(\varepsilon) = (e^{\beta\varepsilon}+1)^{-1}$:
\begin{widetext}
\begin{align}
    \sigma_{CSE}
    &\approx \frac{\partial}{\partial M^2}\frac{\partial}{\partial \mu} \frac{N_c}{\pi^2}  \sum_{n=1}^{+\infty}(-1)^{n}
        \int_0^{+\infty}{p^2dp}
        \left(e^{\beta n(\mu - V_1/2-\sqrt{p^2 + M^2})} - (\mu\to-\mu)\right)\nonumber\\
    &= -\frac{\partial}{\partial M^2}\frac{\partial}{\partial \mu}\frac{N_c}{\pi^2}
        \int_0^{+\infty}{p^2dp}
        \left(f_\beta({\cal E}_M(p) + V_1/2 -\mu ) - f_\beta(\mu\to-\mu)\right)\nonumber\\
        &= \frac{\partial^2}{\partial \mu^2}\frac{N_c}{2\pi^2}
        \int_0^{+\infty}\frac{p^2dp}{\sqrt{p^2 + M^2}}
        \left(f_\beta({\cal E}_M(p) + V_1/2 -\mu ) + f_\beta(\mu\to-\mu)\right)\label{EqICSE0}
\end{align}
\end{widetext}
Here ${\cal E}_M(p)= \sqrt{p^2+M^2} $. One can easily see that in the limiting case $\mu \gg T$:
\begin{align}
	\sigma_{CSE}
&= \frac{\partial}{\partial M^2}\frac{N_c}{\pi^2}
	\int_0^{+\infty}{p^2dp}
	\delta({\cal E}_M(p) + V_1/2 -\mu ) = \frac{N_c}{2\pi^2}\label{EqICSE}
\end{align}
as expected. The dependence of Eq. (\ref{EqICSE0}) on $\mu$ is represented in Fig. \ref{SigmaMu}. It is worth mentioning that strictly speaking at $\mu \gg T_c$ the above expressions for $V_1$ and $M$ cannot be applied. As expected both these quantities decrease essentially with increase of $\mu$. As a result, the value of $\sigma_{CSE}$, in fact, approaches the conventional value faster than represented in Fig. \ref{SigmaMu}.

In Fig. \ref{comparison} we compare our results with those obtained using lattice numerical simulations. 
One can see that qualitatively the two methods give similar results. 
Quantitative difference may be caused by several factors. 
First of all, the finite volume effects may be strong for the lattice simulations with given lattice sizes. 
\red{Next, perturbative corrections disregarded in our calculations may change the results. % by an amount of about $30$  percent.
Also, our result heavily depends on an `unstable' numerical input $V_1(T)$ \eqref{EqV1fit}.}\mzz{ One can see that at small temperatures the FCM gives results that match lattice data if thermal mass is neglected, while at large temperatures the FCM matches lattice results if thermal mass is taken into account while Polyakov line contribution is neglected. The complete FCM interpolates between the two.}

\begin{figure}[h]
	\centering  %
	\includegraphics[width=0.9\linewidth]{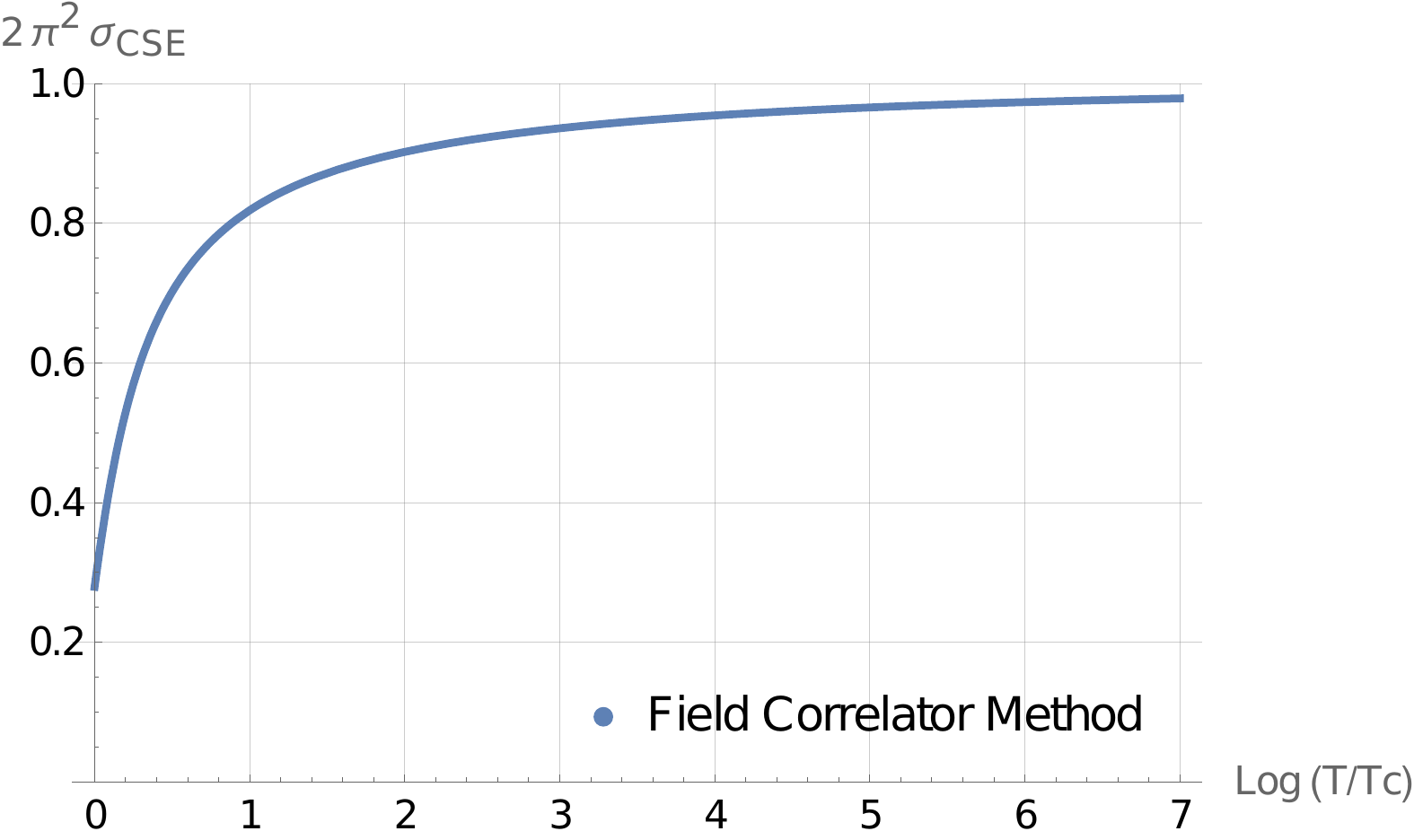}  %
	\caption{We represent here the dependence of $2 \pi \sigma_{CSE}$ (per Dirac fermion, i.e divided by the number of quark flavors $N_f$ and colors $N_c$) \eqref{EqICSE0} at $\mu = T_c$ as a function of $T$. }  %
	\label{SigmaT}   %
\end{figure}

Looking at the results presented in Fig. \ref{SigmaMu} we conclude that at the values of quark chemical potential accessed at LHC, RHIC, NICA, and FAIR the topological regime is not yet achieved, and the CSE conductivity is suppressed essentially compared to the standard topological value.

In addition in Fig. \ref{SigmaT} we represent the dependence of $\sigma_{CSE}$ in units of $\frac{N_c N_f}{2 \pi^2}$ at $\mu = T_c$ as a function of temperature. One can see that this value approaches the conventional one only at the electroweak scale $T \sim 100$ GeV.
\begin{figure}[h]
	\centering  %
	\includegraphics[width=0.9\linewidth]{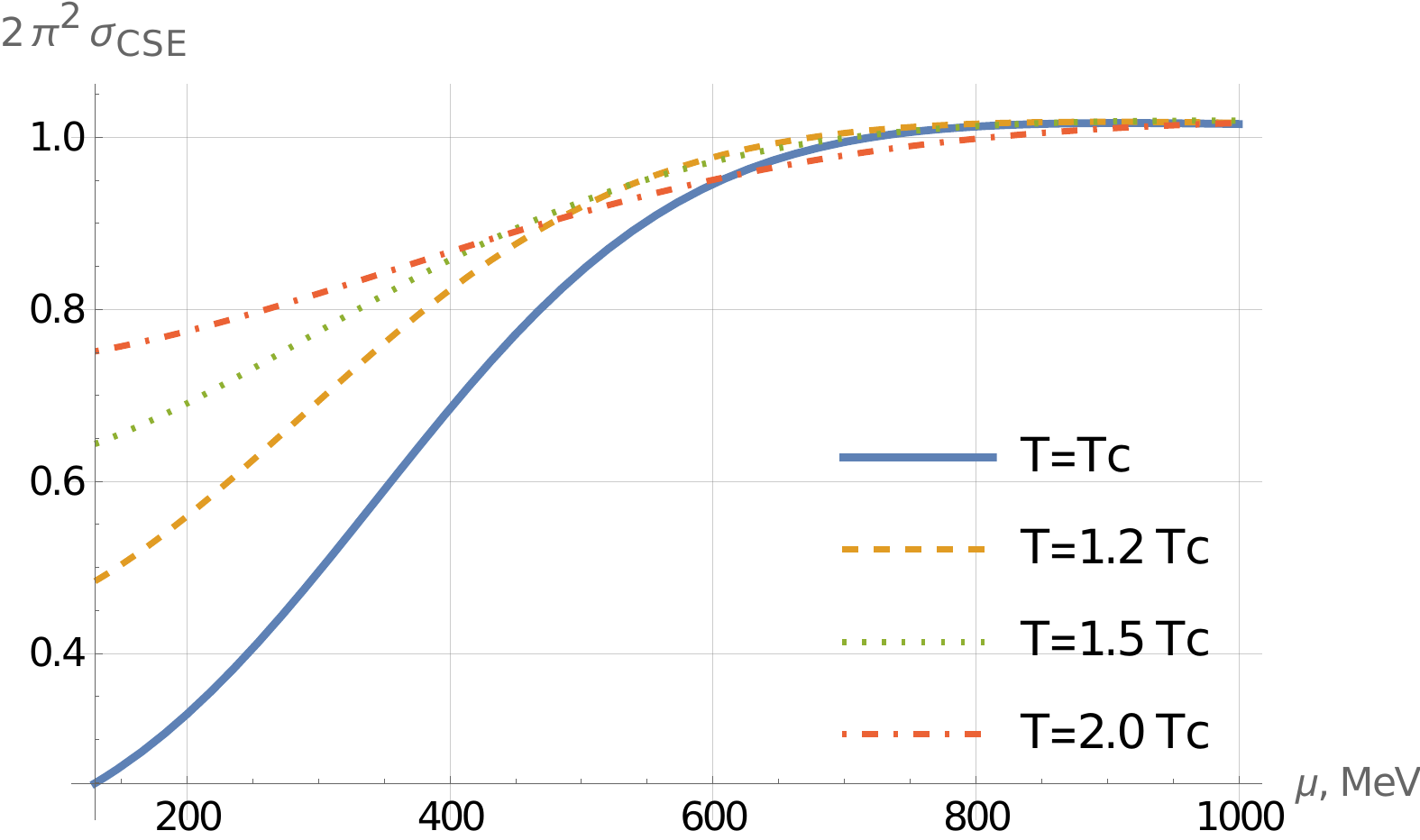}  %
    \caption{We represent here our data on  $2 \pi \sigma_{CSE}$ (per Dirac fermion) \eqref{EqICSE0} as a function of $\mu$ \red{at various temperatures}.  
    }  
	\label{SigmaMu}   %
\end{figure}

\section{Conclusions and discussion}

In the present paper we consider effect of interactions on the chiral separation effect. First of all, we are interested in the effect of strong interactions on the CSE in quark matter. However, the obtained zero temperature results may be applied directly to the CSE in Weyl semimetals as well.

We prove that in the fermion system with chiral symmetry {\it at zero temperature} in the presence of external magnetic field strength $F_{ij}$ and chemical potential $\mu$ the derivative of the \mz{\it renormalized} axial current averaged over the overall volume is given by
\begin{equation}
	\frac{d\bar{J}_5^k}{d\mu} = \frac{\mathcal N}{4\pi^2}\epsilon^{ijk0}  F_{ij}\label{1_f}
\end{equation}
\mz{(By renormalized current we understand expression with bare velocity operator substituted by the renormalized one.) }
This expression is valid provided that $\lambda^2|F_{ij}| \ll 1$, where $\lambda$ is the correlation length of the given system. In particular, for homogeneous cold quark matter with $\mu > \Lambda_{QCD}$ we need magnetic field strength much smaller than $\Lambda_{QCD}^2$.

Here $\cal N$ is the topological invariant given by
\begin{widetext}
	\begin{eqnarray}
		\mathcal{N}&=&\frac{1}{48 \pi^2 {\bf V}}
		\int_{\Sigma_3}
		\int d^3x
		\tr \Bigg[\gamma^5
		{\bf G}_W^{(0)}\star d {\bf Q}_W^{(0)} \star {\bf G}_W^{(0)}
		\wedge \star d {\bf Q}_W^{(0)}\star {\bf G}_W^{(0)} \star \wedge d {\bf Q}_W^{(0)}
		\Bigg]\label{Ncompl_f}
	\end{eqnarray}
\end{widetext}
In this expression \mzo{$\Sigma_3$ is the hypersurface in momentum space consisting of the two hyperplanes $p_4 = \pm \epsilon \to 0$.} $\hat{\bf G}^{(0)}$ is the {\it renormalized} complete two - point Green function with interaction corrections included. \mzz{It has to be calculated after a certain gauge is fixed (if we are speaking of the quark matter). The result does not depend on the chosen gauge.}  Correspondingly, $\hat{\bf Q}^{(0)}$ is operator inverse to $\hat{\bf G}^{(0)}$, while ${\bf Q}^{(0)}_W$ is its Weyl symbol.  Superscript $^{(0)}$ means that the expression for $\hat{\bf G}$ does not contain the external magnetic field. \mzo{Eq. (\ref{Ncompl_f}) may also be represented in the form with integration along the hypersurface $\Sigma_3$ of other forms (but still surrounding the singularities of the expression standing in the integral). For the details see footnote after Eq.  (\ref{eq47}). } 

\mzo{For any $x$ the position of the singularities of an expression standing in the integral inside Eq. (\ref{Ncompl_f}) generalizes the notion of Fermi surface and reduces to it in the homogeneous case. It is supposed that around this generalized Fermi surface matrix $\gamma^5$ commutes (or anti - commutes) with ${\bf Q}^{(0)}_W$.}    This means, actually, that in the given system  at low energies there is chiral symmetry. 

In the system with $N$ chiral Dirac fermions ${\cal N} = N$, and the above result means that the CSE conductivity is given by
$$
\sigma_{CSE} = \frac{N}{2\pi^2}
$$

Being applied to the quark - gluon matter this means that if the dense cold quark matter exists in the phase with restored chiral symmetry without color superconductivity (\mz{we also neglect effect of instantons}), then in this phase the chiral separation effect is present with the conventional expression for the CSE conductivity. The same refers to the quark gluon plasma phase provided that $\mu \gg T$. We calculate directly the non - perturbative corrections to the CSE conductivity at finite temperature above the deconfinement phase transition. Our results confirm that $\sigma_{CSE}$ approaches the conventional expression at large $\mu$ for any given $T>T_c$. However, in the region of the phase diagram accessible at the modern colliders the topological expression for $\sigma_{CSE}$ is not yet approached, and the conductivity is suppressed essentially. It is worth mentioning that at $\mu, T \sim T_c$ the perturbative corrections will give contributions of the order of $30$ percent since  $\alpha_s(T_c) \approx 0.3$. At the same time for large $\mu$ the perturbative corrections are already not so relevant because $\alpha_s$ decreases with the increase of $\mu$. Nevertheless the calculation of the perturbative corrections to $\sigma_{CSE}$ is worth to be performed. But this is out of the scope of the present paper.

The obtained results may also be applied to the CSE in Weyl semimetals, where electronic quasiparticles are subject to Coulomb interactions. Those interactions are typically strong because effective finite structure constant is of the order of unity. Left and right - handed fermions in momentum space are separated here in momentum space. Boundary of the samples contain Fermi arcs. In the presence of magnetic field and chemical potential that exceeds the level of Fermi points the axial current appears. Then the left and the right - handed electrons move in opposite directions. As a result at the boundary of the sample there will be excess of the electrons at the left - handed Weyl point and deficiency of the electrons at the right handed Weyl point (or vice versa). This results in the appearance of the Fermi pockets instead of the Fermi arcs (electron Fermi pocket close to one of the Weyl points, and hole pocket close to the other Weyl point). This is how the CSE effect may be observed experimentally in these materials.

It is worth mentioning that the perturbative calculation of corrections to the CSE conductivity in pure QED performed in  \cite{Shovkovy} suggests the appearance of correction proportional to fine structure constant $\alpha \approx 1/137$, and containing the infrared divergencies. Our approach may be applied effectively to pure QED as well. The essential difference between the two approaches is that in the present paper from the very beginning the renormalized axial current is calculated according to Sect. \ref{SectRenorm}. This approach takes into account the renormalization procedure both for the propagators and for the interaction vertices automatically.  In \cite{Shovkovy} corrections to bare axial current are calculated. Presumably, the  renormalization procedure applied to the expression given in \cite{Shovkovy} will remove completely radiative corrections to the CSE conductivity.

The cousin of the CSE - the chiral vortical effect (CVE) is expected in quark matter under the same conditions as the CSE, i.e. in the same region of the QCD phase diagram. The rotating fireballs containing quark - gluon plasma appear during the non - central heavy ion collisions. The interior of neutron stars may contain quark matter in the phase with restored chiral symmetry. Above we mentioned that this phase might exist without color superconductivity. Rotation results in the appearance of axial current along the axis of rotation in quark systems. At zero temperature  rotation may effectively be described by the effective Abelian gauge field $\mu u_k$, where $u_k$ is the four  vector of rotation velocity \cite{AKZ2018}. As a result   the CVE is reduced to CSE, and the axial current is given by
$$
J_{5} = \frac{{\cal N}}{2\pi^2} \mu^2 \Omega
$$
Here $\Omega$ is angular velocity while
$\cal N$ is given by Eq. (\ref{Ncompl_f}). In cold quark matter ${\cal N} = N_c N_f$, where $N_f$ is the number of quarks with masses smaller than $\mu$. The same expression might also be applied to quark gluon plasma at $\mu \gg T$ if the rotation is considered as rigid. Actually, the fireballs do not rotate rigidly. Therefore, this approach may be taken into account only qualitatively. In the domain $\mu \sim T$ the mentioned methodology, in which rotation is introduced through the effective Abelian gauge field, cannot be applied at all. Namely, rotation of thermal quasiparticles cannot be described by effective Abelian gauge field. Rotating thermal quasiparticles contribute the axial current along the axis of rotation. These contributions should be taken into account separately. \mzz{Notice that in \cite{Hou_2012, Golkar_2015} it has been pointed out that the temperature depending term in the CVE conductivity does receive interaction corrections resulted from the exchange by gauge bosons. However, in \cite{Hou_2012} it was argued that this term is not subject to corrections resulted from Yukawa interactions.}

\begin{appendices} \begin{section}{}\label{App1}

		We present here the calculation of quark propagator using method of the main text applied to the non - interacting system at finite temperature. This way we check the normalization factor entering measure in the path integral over trajectories \eqref{EqPathInt}:
		\begin{widetext}
		\begin{align}
		-	S(x, y) =& (\slashed{\partial} - m)G(x,y), \\
			G(x, y) =&
			\int_0^{+\infty}ds(\overline{\cD^4z})_{xy}^s \exp\left(-m^2s - \frac14\int_0^s\dot{z}^2d\tau\right) \\
			=& \int_0^{+\infty}ds
			\lim_{\mathbin{\ensurestackMath{\abovebaseline%[-3.4pt]
						{\stackunder%[-3.5pt]
							{\scriptstyle{N\to +\infty}}{\scriptstyle{\varepsilon=s/N}}}}}}
			\left(\prod_{m=1}^{N}\frac{d^4\D z_m}{(4\pi\varepsilon)^2}\right)
			\sum_{n=-\infty}^{+\infty}(-1)^n \frac{d^4p}{(2\pi)^4} \times\nonumber\\
			&\times\exp\left(ip_\mu\left(\sum_{i=1}^{N}\D z^\mu_i - (x-y)^\mu - n\beta\delta^\mu_4\right)
			- m^2s - \sum_{i=1}^{N}\frac{\D z^2_i}{4\varepsilon}\right) \\
			=& \int_0^{+\infty}ds \sum_{n=-\infty}^{+\infty}(-1)^n\int \frac{d^4p}{(2\pi)^4}
			\exp\left(-(p^2+m^2)s - ip\cdot(x-y) - ip_4 n\beta\right) \\
			=& \sum_{n=-\infty}^{+\infty}(-1)^n\int \frac{d^4p}{(2\pi)^4}
			\frac{\exp\left(- ip\cdot(x-y) - ip_4 n\beta\right)}{p^2+m^2}
		\end{align}
	\end{widetext}
		The anti-periodic boundary conditions in imaginary time are assumed here \cite{LeBellac2011kqa}
		\begin{equation}
			(\slashed{\partial} + m)S(x) = \delta(x),
			\quad S(\tau+\beta, \vec x) = -S(\tau, \vec x).
		\end{equation}
			In coordinate representation the propagator reads
		\begin{eqnarray}
			G(x, y) &=& \sum_{n=-\infty}^{+\infty}(-1)^n \int_0^{+\infty}
			\frac{ds }{(4\pi s)^2}\exp\left(-m^2s-\frac{z^2_n}{4s}\right)
		\end{eqnarray}
	This gives
		\begin{eqnarray}
		G(x, y) &=& \sum_{n=-\infty}^{+\infty}(-1)^n \frac{m}{4\pi^2z_n}K_1(mz_n),
	\end{eqnarray}
	where $z_n^2 = (\vec x- \vec y)^2+(x_4-y_4+n\beta)^2$.
		
\end{section}

\end{appendices}

\bibliographystyle{utphys}
\bibliography{CSE_MZ.bib}

\end{document}